\documentclass[11pt,a4paper]{article}
\pdfoutput=1

\usepackage{jheppub}
\usepackage[utf8]{inputenc}
\usepackage{xspace}
\usepackage{url}

\usepackage{verbatim}
\usepackage{amsthm,mathtools}
\usepackage{relsize}
\usepackage[mathscr]{euscript}
\usepackage[cal=dutchcal]{mathalfa}
\pdfmapfile{+rsfso.map}
\DeclareSymbolFont{rsfso}{U}{rsfso}{m}{n}
\DeclareSymbolFontAlphabet{\mathscr}{rsfso}

\usepackage{float}
\usepackage[section]{placeins}
\usepackage[lofdepth,lotdepth]{subfig}
\usepackage[toc,page]{appendix}
\usepackage[dvipsnames]{xcolor}
\usepackage{xtab}
\usepackage{enumitem}
\usepackage[hhmmss]{datetime}

\usepackage{tikz}
\usetikzlibrary{shapes,arrows}

\usepackage{booktabs}
\allowdisplaybreaks

\usepackage{tabu}
\usepackage{booktabs}
\usepackage{multirow}
\newcolumntype{P}[1]{>{\centering\arraybackslash}p{#1}}

\hypersetup{hidelinks}

\makeatletter

\AtBeginDocument{%
	\expandafter\renewcommand\expandafter\subsection\expandafter{%
		\expandafter\@fb@secFB\subsection
	}%
}

\makeatother % this adds \FloatBarrier at the end of subsections, so floating objects are obliged to stay in the subsection you want them to stay

\newcommand{\rb}[1]{\left(#1\right)}
\newcommand{\qb}[1]{\left[#1\right]}
\newcommand{\cb}[1]{\left\lbrace #1 \right\rbrace}

\newcommand{\mydet}[1]{\operatorname{det}\left({#1}\right)}

\newif\ifColors    \Colorsfalse
\newif\ifShowSigns    \ShowSignstrue

\hypersetup{
	pdfstartview={Fit}, pdfpagelayout={SinglePage}
}

\newcommand{\change}[1]{{#1}}

\usepackage{mathabx}
\usepackage{accents}
\usepackage{xspace}

\DeclareMathAlphabet\urwscr{U}{urwchancal}{m}{n}%

\ifx \ii \undefined  % Avoid the redefinitions
\DeclareMathAlphabet{\mathsfit}{\encodingdefault}{\sfdefault}{m}{sl}
\SetMathAlphabet{\mathsfit}{bold}{\encodingdefault}{\sfdefault}{bx}{sl}

\DeclareSymbolFont{eulerletters}{U}{eus}{m}{n}
\DeclareMathSymbol{\eulU}{\mathord}{eulerletters}{`U}
\DeclareMathSymbol{\eulC}{\mathord}{eulerletters}{`C}
\DeclareMathSymbol{\eulB}{\mathord}{eulerletters}{`B}
\DeclareMathSymbol{\eulQ}{\mathord}{eulerletters}{`Q}
\DeclareMathSymbol{\eulE}{\mathord}{eulerletters}{`E}
\DeclareMathSymbol{\eulD}{\mathord}{eulerletters}{`D}
\DeclareMathSymbol{\eulV}{\mathord}{eulerletters}{`V}
\DeclareMathSymbol{\eulW}{\mathord}{eulerletters}{`W}
\DeclareMathSymbol{\eulS}{\mathord}{eulerletters}{`S}
\DeclareMathSymbol{\eulG}{\mathord}{eulerletters}{`G}
\DeclareMathSymbol{\eulL}{\mathord}{eulerletters}{`L}
\DeclareMathSymbol{\eulK}{\mathord}{eulerletters}{`K}
\DeclareMathSymbol{\eulR}{\mathord}{eulerletters}{`R}

\DeclareSymbolFont{CMB}{OMS}{cmsy}{m}{it}
\DeclareMathSymbol{\cmA}{\mathord}{CMB}{`A}
\DeclareMathSymbol{\cmB}{\mathord}{CMB}{`B}
\DeclareMathSymbol{\cmW}{\mathord}{CMB}{`W}

\definecolor{red}{rgb}{1,0,0.1}
\definecolor{green}{rgb}{0.0,0.6,0}
\definecolor{blue}{rgb}{0.1,0.1,1}
\definecolor{orange}{rgb}{0.6,0.3,0}
\definecolor{magenta}{rgb}{0.9,0.1,1}

\newcommand{\qm}[1]{``#1"}

\newcommand\nPlusOne{$\sdim\hspace{-0.1em}+\hspace{-0.1em}1$\xspace}
\newcommand\threePlusOne{$3\hspace{-0.075em}+\hspace{-0.1em}1$\xspace}
\newcommand\onePlusLog{1+$\hspace{0.1em}\log$\xspace}

\newcommand\transpose[1]{{#1}^\mathsmaller{\mathsf{T}}}

\renewcommand\tilde[1]{\mkern1mu\widetilde{\mkern-1mu#1}}

\newcommand\mat[1]{
\begin{pmatrix}
#1
\end{pmatrix}
}

\newcommand\diagmat[1]{\operatorname{diag}\qb{#1}}

\usepackage{xparse}
\ExplSyntaxOn
\NewDocumentCommand{\mref}{m}{\quinn_mref:n {#1}}
\seq_new:N \l_quinn_mref_seq
\cs_new:Npn \quinn_mref:n #1
 {
  \seq_set_split:Nnn \l_quinn_mref_seq { , } { #1 }
  \seq_pop_right:NN \l_quinn_mref_seq \l_tmpa_tl
  ( % print the left parenthesis
  \seq_map_inline:Nn \l_quinn_mref_seq
    { \ref{##1},\nobreakspace } % print the first references
  \exp_args:NV \ref \l_tmpa_tl % print the last or only one
  ) % print the right parenthesis
 }
\ExplSyntaxOff

\newcommand\ii{\mathrm{i}}
\newcommand\ee{\mathrm{e}}

\newcommand\tr{\mathsf{{\scriptscriptstyle T}}}

\newcommand\dt{\partial _t}
\newcommand\dr{\partial _r}
\newcommand\drr{\partial _{rr}}

\newcommand\stdim{d}
\newcommand\sdim{N}

\ifColors
  \newcommand\gSector[1]{{\color{blue}#1}}
  \newcommand\fSector[1]{{\color{red}#1}}
  \newcommand\hSector[1]{{\color{green}#1}}
  \newcommand\lSector[1]{{\color{BurntOrange}#1}}
  \newcommand\mSector[1]{{\color{magenta}#1}}

  \newcommand\betaSum{\,{\color{orange}m^{\stdim}{\sum_{n=0}^\stdim}\beta_{n}}}
  
\else
  \newcommand\gSector[1]{{\color{black}#1}}
  \newcommand\fSector[1]{{\color{black}#1}}
  \newcommand\hSector[1]{{\color{black}#1}}
  \newcommand\lSector[1]{{\color{black}#1}}
  \newcommand\mSector[1]{{\color{black}#1}}

  \newcommand\betaSum{\,{m^{\stdim}{\sum_{n=0}^\stdim}\beta_{n}}}
  
\fi

\newcommand\gMet{\gSector g}
\newcommand\gSp{\gSector{\gamma}}
\newcommand\gLapse{\gSector{\alpha}}
\newcommand\gShift{\gSector \beta}

\newcommand\gK{\gSector K}
\newcommand\gE{\gSector e}

\newcommand\gD{\gSector{D}}
\newcommand\gR{\gSector R}

\newcommand\gVse{\gSector{V_{g}}}
\newcommand\gTse{\gSector{T_{g}}}
\newcommand\gTeff{\gSector{T_{g}^\mathrm{eff}}}
\newcommand\gEinst{\gSector{G_{g}}}

\newcommand\gCC{\gSector{\eulC}}

\newcommand\fMet{\fSector f}
\newcommand\fSp{\fSector{\varphi}}
\newcommand\fLapse{\fSector{\tilde{\alpha}}}
\newcommand\fShift{\fSector{\tilde{\beta}}}

\newcommand\fK{\fSector{\tilde{K}}}
\newcommand\fE{\fSector m}
\newcommand\fEb{\fSector{m_\mathsf{1}}}
\newcommand\fEtr{\fSector{m_\mathsf{o}}}

\newcommand\fD{\fSector{\tilde{D}}}
\newcommand\fR{\fSector{\tilde{R}}}

\newcommand\fVse{\fSector{V_{f}}}
\newcommand\fTse{\fSector{T_{f}}}
\newcommand\fTeff{\fSector{T_{f}^\mathrm{eff}}}
\newcommand\fEinst{\fSector{G_{f}}}

\newcommand\fCC{\fSector{\widetilde{\eulC}}}

\newcommand\mS{\mSector{\eulC_\mathrm{b}}}

\newcommand\mW{\mSector{W}}

\newcommand\gKappa{\gSector{\kappa_{g}}}

\newcommand\fKappa{\fSector{\kappa_{f}}}

\newcommand\grho{\gSector{\rho^\mathrm{m}}}
\newcommand\gjota{\gSector{j^\mathrm{m}}}
\newcommand\gJota{\gSector{J^\mathrm{m}}}

\newcommand\frho{\fSector{\tilde{\rho}^\mathrm{m}}}
\newcommand\fjota{\fSector{\tilde{j}^\mathrm{m}}}
\newcommand\fJota{\fSector{\tilde{J}^\mathrm{m}}}

\newcommand\grhob{\gSector{\rho^\mathrm{b}}}
\newcommand\gjotab{\gSector{j^\mathrm{b}}}
\newcommand\gJotab{\gSector{J^\mathrm{b}}}

\newcommand\frhob{\fSector{\tilde{\rho}{}^\mathrm{b}}}
\newcommand\fjotab{\fSector{\tilde{j}{}^\mathrm{b}}}
\newcommand\fJotab{\fSector{\tilde{J}{}^\mathrm{b}}}

\newcommand\grhoeff{\gSector{\rho^\mathrm{eff}}}
\newcommand\gjotaeff{\gSector{j^\mathrm{eff}}}
\newcommand\gJotaeff{\gSector{J^\mathrm{eff}}}

\newcommand\frhoeff{\fSector{\tilde{\rho}{}^\mathrm{eff}}}
\newcommand\fjotaeff{\fSector{\tilde{j}{}^\mathrm{eff}}}
\newcommand\fJotaeff{\fSector{\tilde{J}{}^\mathrm{eff}}}

\newcommand\gEA{\gSector A}

\newcommand\fEA{\fSector{\tilde{A}}}

\newcommand\sgn{\gSector{\mathsfit{n}{\mkern1mu}}}
\newcommand\sgD{\gSector{\eulD}}
\newcommand\sgQ{\gSector{\eulQ}}
\newcommand\sgV{\gSector{\eulV}}
\newcommand\sgU{\gSector{\eulU}}
\newcommand\sgB{\gSector{\eulB}}

\newcommand\sfn{\fSector{\tilde{\mathsfit{n}}{\mkern1mu}}}
\newcommand\sfD{\fSector{\widetilde{\eulD}}}
\newcommand\sfQ{\fSector{\widetilde{\eulQ}}}
\newcommand\sfV{\fSector{\widetilde{\eulV}}}
\newcommand\sfU{\fSector{\widetilde{\eulU}}}
\newcommand\sfB{\fSector{\widetilde{\eulB}}}

\newcommand\sgW{\gSector{W_g}}%\cmW
\newcommand\sgQU{\gSector{(\eulQ\fSector{{\scriptstyle \widetilde{\eulU}}})}}

\newcommand\sfW{\fSector{W_f}}%\cmW
\newcommand\sfQU{\fSector{(\widetilde{\eulQ}\gSector{{\scriptstyle \eulU}})}}

\newcommand\gBSSN[1]{{\widebar{#1}}}
\newcommand\fBSSN[1]{{\widehat{#1}}}
\newcommand\hBSSN[1]{{\accentset{\circ}{{#1}}}}
\newcommand\lBSSN[1]{{\accentset{\ast}{{#1}}}}

\newcommand\hSpBS{\hSector{\hBSSN \chi}}

\newcommand\pfg{\gSector{\partial_{\perp}}}
\newcommand\pff{\fSector{\widetilde{\partial}_{\perp}}}

\newcommand\gconf{\gSector{\phi}}
\newcommand\fconf{\fSector{\psi}}

\newcommand\gSpBS{\gSector{\gBSSN \gamma}}

\newcommand\gShiftB{\gSector B}

\newcommand\gABS{\gSector{\gBSSN A}}
\newcommand\gKBS{\gSector{\gBSSN K}}
\newcommand\gEBS{\gSector{\gBSSN e}}
\newcommand\gDBS{\gSector{\gBSSN{D}}}
\newcommand\gCCBS{\gSector{\gBSSN{\eulC}}}
\newcommand\gCGBS{\gSector{\gBSSN{\eulG}}}

\newcommand\gGBS{\gSector{\gBSSN \Gamma}}
\newcommand\gL{\gSector{\gBSSN \Lambda}}
\newcommand\gDG{\gSector{\bigtriangleup\gBSSN{\Gamma}}}
\newcommand\gDback{\gSector{\gBSSN{D}_\mathsf{B}}}
\newcommand\gGbackBS{\gSector{\gBSSN{\Gamma}_\mathsf{B}}}
\newcommand\gRback{\gSector{\gBSSN{R}_\mathsf{B}}}
\newcommand\gRBS{\gSector{\gBSSN{\eulR}}}
\newcommand\gRiBS{\gSector{\gBSSN{R}}}

\newcommand\gA{\gSector{\gBSSN{a}}}
\newcommand\gB{\gSector{\gBSSN{b}}}
\newcommand\gAo{\gSector{\gBSSN{A}_1}}
\newcommand\gAt{\gSector{\gBSSN{A}_2}}
\newcommand\gLo{\gSector{\gBSSN{\Lambda}^r}}

\newcommand\po{\lSector{\sLp^\textbf{1}}}
\newcommand\pt{\lSector{\sLp^\textbf{2}}}
\newcommand\pth{\lSector{\sLp^\textbf{3}}}
\newcommand\qo{\hSector{\hShift^r}}
\newcommand\qt{\hSector{\hShift^\theta}}
\newcommand\qth{\hSector{\hShift^\phi}}

\newcommand\fSpBS{\fSector{\fBSSN \varphi}}

\newcommand\fABS{\fSector{\fBSSN A}}
\newcommand\fKBS{\fSector{\fBSSN K}}
\newcommand\fEBS{\fSector{\fBSSN m}}
\newcommand\fEbBS{\fSector{\fBSSN{m}_\mathtt{1}}}
\newcommand\fEtrBS{\fSector{\fBSSN{m}_\mathrm{o}}}
\newcommand\fDBS{\fSector{\fBSSN{D}}}
\newcommand\fCCBS{\fSector{\fBSSN{\eulC}}}
\newcommand\fCGBS{\fSector{\fBSSN{\eulG}}}

\newcommand\fGBS{\fSector{\fBSSN \Gamma}}
\newcommand\fL{\fSector{\fBSSN \Lambda}}
\newcommand\fDG{\fSector{\bigtriangleup\fBSSN{\Gamma}}}
\newcommand\fDback{\fSector{\fBSSN{D}_\mathsf{B}}}
\newcommand\fGbackBS{\fSector{\fBSSN{\Gamma}_\mathsf{B}}}
\newcommand\fRback{\fSector{\fBSSN{R}_\mathsf{B}}}
\newcommand\fRBS{\fSector{\fBSSN{\eulR}}}
\newcommand\fRiBS{\fSector{\fBSSN{R}}}

\newcommand\fA{\fSector{\fBSSN{a}}}
\newcommand\fB{\fSector{\fBSSN{b}}}
\newcommand\fAo{\fSector{\fBSSN{A}_1}}
\newcommand\fAt{\fSector{\fBSSN{A}_2}}
\newcommand\fLo{\fSector{\fBSSN{\Lambda}^r}}

\newcommand\gdetBS{\gSector{\gBSSN \Delta}}
\newcommand\fdetBS{\fSector{\fBSSN \Delta}}
\newcommand\hdetBS{\hSector{\accentset{\circ}{\Delta}}}

\newcommand\sgnBS{\gSector{\gBSSN{\mathsfit{n}{\mkern1mu}}}}
\newcommand\sgDBS{\gSector{\gBSSN \eulD}}
\newcommand\sgQBS{\gSector{\gBSSN \eulQ}}

\newcommand\sgBBS{\gSector{\gBSSN \eulB}}

\newcommand\sfnBS{\fSector{\fBSSN{\mathsfit{n}{\mkern1mu}}}}
\newcommand\sfDBS{\fSector{\fBSSN \eulD}}
\newcommand\sfQBS{\fSector{\fBSSN \eulQ}}

\newcommand\sfBBS{\fSector{\fBSSN \eulB}}

\newcommand\mSqrt{\mSector{S}}

\newcommand\sLsBS{\lSector{\boldsymbol{\lBSSN{\Lambda}}_\mathrm{s}}}
\newcommand\sRsBS{\lSector{\boldsymbol{\lBSSN{\mathrm{R}}}}}
\newcommand\sRbarBS{\lSector{\boldsymbol{\lBSSN{\mathrm{R}}_\mathrm{o}}}}
\newcommand\sLtBS{\lSector{\lBSSN{\lambda}}}
\newcommand\sLvBS{\lSector{\lBSSN{\boldsymbol{\mathrm{v}}}}}
\newcommand\sLpBS{\lSector{\lBSSN{\boldsymbol{\mathrm{p}}}}}
\newcommand\svec{\lSector{\boldsymbol{\mathrm{u}}}}

\newcommand\hMet{\hSector h}
\newcommand\hSp{\hSector{\chi}}
\newcommand\hLapse{\hSector H}
\newcommand\hShift{\hSector q}

\newcommand\sLs{\lSector{\boldsymbol{\Lambda}_\mathrm{s}}}
\newcommand\sLt{\lSector{\lambda}}

\newcommand\sLv{\lSector{\boldsymbol{\mathrm{v}}}}
\newcommand\sLp{\lSector{\boldsymbol{\mathrm{p}}}}
\newcommand\sRs{\lSector{\boldsymbol{\mathrm{R}}}}
\newcommand\sRbar{\lSector{\boldsymbol{\mathrm{R}}_\mathrm{o}}}

\newcommand\sI{\lSector{\boldsymbol{\mathrm{I}}}}
\newcommand\sEta{\lSector{\boldsymbol{\mathrm{\delta}}}}

\newcommand\gdet{\gSector{\Delta}}
\newcommand\fdet{\fSector{\tilde \Delta}}
\newcommand\hdet{\hSector{\overset{\scriptscriptstyle{\#}}{\Delta}}}

\newcommand\mR{\mSector{\wideparen{R}}}
\newcommand\esp[2]{\left\langle\mR\right\rangle ^{#1}_{#2}}

\ifColors

\else

\fi

\fi  % avoid the redefinitions
\title{Covariant BSSN formulation in bimetric relativity}

\author{Francesco Torsello,}
\emailAdd{francesco.torsello@fysik.su.se}

\author{Mikica Kocic,}
\author{Marcus H\"ogås,}
\author{Edvard M\"ortsell}

\affiliation{Department of Physics \& The Oskar Klein Centre, \\
Stockholm University, AlbaNova University Center, SE-106 91 Stockholm, Sweden}

\abstract{Numerical integration of the field equations in bimetric relativity is necessary to obtain solutions describing realistic systems. Thus, it is crucial to recast the equations \change{as a well-posed problem}. In general relativity, under certain assumptions, the covariant BSSN formulation is a \change{strongly hyperbolic} formulation of the Einstein equations, \change{hence its Cauchy problem is well-posed}. In this paper, we establish the covariant BSSN formulation of the bimetric field equations. It shares many features with the corresponding formulation in general relativity, but there are a few fundamental differences between them. Some of these differences depend on the gauge choice and alter the hyperbolic structure of the system of partial differential equations compared to general relativity. Accordingly, the strong hyperbolicity of the system cannot be claimed yet, under the same assumptions as in general relativity. In the paper, we stress the differences compared with general relativity and state the main issues that should be tackled next, to draw a road map towards numerical bimetric relativity.}


\begin{document}

\maketitle

\section{Introduction}

The Hassan--Rosen bimetric theory, or bimetric relativity (BR), is an extension of general relativity (GR) describing the nonlinear interaction of two metrics $\gMet_{\mu\nu}$ and $\fMet_{\mu\nu}$ \cite{Hassan:2011zd,Hassan:2011ea}. Its unambiguous definition and spacetime interpretation are established in \cite{HassanKocic2018}, and the complete Hamiltonian analysis of it is performed in \cite{HassanLundkvist2018}. \change{The study of this theory is well-motivated. First, its cosmology} is compatible with local gravity tests, and in particular with the recent observations of gravitational waves \cite{10.1088/1361-6382/ab4f9b}, since it describes the dynamics of both a massless and a massive spin-2 field, allowing for gravitational waves propagating with the speed of light. In addition, BR provides us with self-accelerating cosmological solutions \cite{Volkov:2011an,vonStrauss:2011mq,Comelli:2011zm,Akrami:2012vf} and a framework where the gravitational origin of dark matter can be studied \cite{Aoki:2014cla,Blanchet:2015bia,Enander:2015kda,Babichev:2016bxi}. This motivates further exploration of the theory and the work to obtain \change{solutions} to its field equations \change{describing realistic physical systems}. These are needed to compare the predictions of the theory against the observational data\change{, possibly confirming or falsifying the theory.}

\change{At first, introducing a second metric in the same spacetime may sound exotic, but turns out to be necessary to make a spin-2 field massive.\footnote{There is no way to construct a dynamical mass term with only one metric, since $\gMet_{\mu\nu}\gMet^{\mu\nu}=4$, the spacetime dimension.} In GR, it is possible to couple a spin-2 field (the metric) to spin-0 and spin-1 fields. Hence, it is natural to try and see if and how a second spin-2 field can be coupled to the metric. Then, the question about the physical meaning of the second metric arises. In BR, independent matter sources are minimally coupled to only one of the two metrics \cite{PhysRevD.90.124042,Rham_2015}. This implies that a test particle coupled to the metric $\gMet_{\mu\nu}$ follows the geodesics determined by this metric, exactly as in GR. The interaction with $\fMet_{\mu\nu}$ is only experienced indirectly. The key difference is that the metric $\gMet_{\mu\nu}$ is not determined by the Einstein field equations (EFE), but rather by the bimetric field equations (BFE) which account for the interaction between the metrics. For more details about BR, we refer the reader to the review in \cite{Schmidt_May_2016}.}

The field equations governing the dynamics of the two metrics have been solved analytically in some important cases (see \cite{Schmidt_May_2016} and references therein). In the majority of those cases, they are reduced to a set of ordinary differential equations. This is done, e.g., by imposing spherical symmetry and staticity---the radial coordinate is the sole independent variable---or spatial homogeneity and spherical symmetry---the time coordinate is the sole independent variable. Other exact solutions are equivalent to those of GR, which one can always obtain in BR in vacuum under the conditions established in \cite{doi:10.1142/S0218271814430020,doi:10.1063/1.5100027}, and in the presence of external matter sources under other conditions \cite{doi:10.1142/S0218271814430020}.

We are interested in solving the bimetric field equations to obtain more realistic solutions, e.g., the spherical gravitational collapse of matter. In this case, very few results can be obtained analytically (see, e.g., \cite{Hogas:2019cpg,Hogas:2019ywm}) and the numerical integration of the BFE becomes indispensable. In addition, the BFE are now a system of partial differential equations (PDEs)---both the time and radial coordinate are independent variables---and, as such, it is desirable to recast them \change{as a well-posed problem} before integrating them. Well-posedness can be naively thought of as the concept of \qm{stability against small perturbations} for PDEs \cite{courant1962methods}. The concept of well-posedness for first-order \change{systems of} PDEs is introduced in \autoref{subsec:wellpos}.

In GR, the EFE can be recasted in a variety of well-posed forms \change{(see, e.g., \cite[Ch.~5]{alcubierre2008introduction})}, one of them being the Baumgarte, Shapiro, Shibata, Nakamura, \change{Oohara and Kojima (BSSNOK or, more commonly, BSSN)} formulation \cite{10.1143/PTPS.90.1,PhysRevD.52.5428,PhysRevD.59.024007,PhysRevD.66.064002,PhysRevD.70.104004} (see also \cite{baumgarte2010numerical} \change{for a pedagogical intorduction}), with its covariant generalization \cite{PhysRevD.79.104029}. One begins by rewriting the EFE as a Cauchy problem. This can be accomplished doing a \nPlusOne decomposition, where the field equations split into a set of constraint equations not involving time derivatives, and a set of evolution equations involving first-order time derivatives.
\change{Then, in the free evolution scheme, one finds the appropriate initial data by solving the constraint equations, and evolves them by solving the evolution equations only. In \cite{Frittelli:1996nj} it is proven that the constraints stay satisfied in the free evolution scheme, and the same holds in BR \cite{Kocic2019con}. If one adopts the free evolution scheme, as we do, only the well-posedness of the Cauchy problem arising from the evolution equations is relevant.}
\begin{comment}
\change{Then, one finds the appropriate initial data by solving the constraint equations, and evolves them by solving the evolution equations. Note that, in principle, one should check that the constraints stay satisfied during the evolution. In \cite{Frittelli:1996nj} is proven that, if one solves the constraints on the initial hypersurface to find the initial data, and then evolves the data by solving the evolution equations only, the constraints stay satisfied. The same holds in BR \cite{Kocic2019con}. This method is called \qm{free evolution scheme} and if one adopts it, as we do, only the well-posedness of the Cauchy problem arising from the evolution equations is relevant.}
\end{comment}

Following the path outlined by numerical relativity, we would like to recast the BFE \change{as a well-posed problem}. The first step, i.e., the \nPlusOne decomposition of the BFE, was established in \cite{Kocic:2018ddp}. In this paper, we present the covariant BSSN (cBSSN) formulation of the BFE. \change{The choice of the BSSN formulation is motivated by the fact that it is one of the most widely used in numerical relativity, allowing for stable long-term numerical evolution---see, e.g., subsection 11.4.5 in \cite{gourgoulhon20123+1}. A future research goal is to obtain stable long-term bimetric simulations, hence, as the starting point, we followed closely what people are doing in numerical relativity. Examples of guiding lights are the Einstein Toolkit \cite{Lffler2012} and SENR/NRPy+ \cite{PhysRevD.97.064036}, which are both using the BSSN formulation.} We show that, even though the cBSSN formulation of the \change{EFE}, together with the standard gauge and some other technical assumptions, is strongly hyperbolic and therefore \change{its Cauchy problem is} well-posed and suitable for the numerical integration \cite{PhysRevD.79.104029}, we cannot yet say if the cBSSN formulation of the BFE is also strongly hyperbolic. The reason is that the lapse functions of the two metrics are dependent \cite{Alexandrov2013,HassanLundkvist2018,Kocic_2019}. Their ratio, after a first-order reduction \change{of the equations}, involves the dynamical fields algebraically.\footnote{The shift vectors are dependent as well, and their relation contains the ratio between the lapses.} This ratio appears in the equations in both \change{the metric} sectors and, depending on the gauge choice, its first and second spatial derivatives can appear in the equations of one or both sectors. These spatial derivatives affect the hyperbolic structure of the system of PDEs compared to GR, hence the result about the \change{strong hyperbolicity} of the cBSSN formulation of the EFE cannot be directly extended to the cBSSN formulation of the BFE. In the paper, we discuss in more detail these issues and stress the similarities and differences compared to GR.  \change{Observe that the bimetric features described in the paper, such as the discussion about the gauge fixing, the relations between lapses and shifts, and how these affect the well-posedness of the problem, do not depend on the chosen formulation and need to be accounted for in any formulation of the BFE.}

We stress that recasting the BFE \change{as a well-posed Cauchy problem, suitable for numerical integration,} is a powerful strategy to obtain interesting physical solutions in BR. For example, since the Birkhoff theorem is not valid in BR \cite{Kocic:2017hve}, a non-static spherically symmetric system, generically, emits gravitational radiation which is longitudinally polarized, coming from the helicity-0 mode of the massive spin-2 field. This could happen during a spherically symmetric gravitational collapse, or for non-static spherically symmetric black holes. This can provide us with predictions from BR that can be tested against the observational data in the future, which is one of the motivations of our work, inspired by the success of numerical relativity.

\paragraph{Structure of the paper.} In \autoref{subsec:wellpos}, we concisely introduce the concepts of well-posedness, and strong hyperbolicity for a system of (first-order) PDEs. In \autoref{sec:3p1}, we very briefly introduce bimetric relativity and review its $\sdim+1$ formulation. In \autoref{sec:BSSN}, we introduce the BSSN formulation with its covariant extension and emphasize the differences between BR and GR. In \autoref{sec:gauge}, we discuss gauge fixing in bimetric relativity and qualitatively describe how it affects the hyperbolic structure of the evolution equations. We summarize our results and state our view about what the next challenges in this field are in \autoref{sec:sum}. The \hyperref[app:abeqs]{appendix} includes explicit equations and technical details.

\paragraph{Notation.} Consider the spatial metrics $\gSp$, $\fSp$ and $\hSp$ and their determinants. We shall denote the determinant of a spatial metric with $\Delta$, and define the following notation referring to the metric sectors,
\begin{alignat}{4}
	\gdet	 		&\ ,\quad &\mbox{no accent} &:\quad &&\mbox{quantity refers to the $\gMet$-sector},& \nonumber \\
	\fdet 			&\ ,\; &\mbox{tilde}&:&& \mbox{quantity refers to the $\fMet$-sector},& \nonumber \\
	\hdet 		&\ ,\; &\mbox{hash}&: &&\mbox{quantity refers to the $\hMet$-sector},& \nonumber \\
	\lSector{\boldsymbol{\Delta}} 		&\ ,\; &\mbox{boldface}&: &&\text{quantity refers to the Lorentz frame}.& \nonumber \\
	\gdetBS 		&\ ,\quad &\mbox{overbar} &:\quad &&\mbox{quantity refers to the $\gMet$-sector in BSSN},& \nonumber \\
	\fdetBS 			&\ ,\; &\mbox{wide hat}&:&& \mbox{quantity refers to the $\fMet$-sector in BSSN},& \nonumber \\
	\hdetBS 		&\ ,\; &\mbox{circle}&: &&\mbox{quantity refers to the $\hMet$-sector in BSSN},& \nonumber \\
	\lSector{\lBSSN{\boldsymbol{\Delta}}} 		&\ ,\; &\mbox{boldface, asterisk}&: &&\mbox{quantity refers to the Lorentz frame in BSSN}.& \nonumber 
\end{alignat}
We denote tensors both with and without their indices, e.g., the metric $\gMet$ or $\gMet _{\mu\nu}$. Greek indices run from 0 to $\stdim-1$, where $\stdim$ is the dimension of spacetime; latin indices run from 1 to $\stdim-1$; boldface indices are spatial Lorentz indices and run from 1 to $\stdim-1$.

\subsection{Well-posedness and strong hyperbolicity}
\label{subsec:wellpos}

When dealing with PDEs, it is important to be able to write \change{the mathematical boundary value problem arising from them} in a well-posed way. \change{A \qm{mathematical boundary value problem} is a differential problem with some specified boundary or initial data, such as the Dirichlet problem or the Cauchy problem.} The definition of a well-posed \change{problem} given in \cite[p.~226]{courant1962methods} was introduced for the first time by Hadamard in \cite{hadamard1902} and reads: ``A mathematical problem which is to correspond to physical reality should satisfy the following basic requirements: (1) The solution must exist. (2) The solution should be uniquely determined. (3) The solution should depend continuously on the data [...]". As stated in \cite[p.~8]{tikhonov1977solutions}, the first two requirements characterize the mathematical determinacy of the problem, whereas the third requirement characterizes its physical determinacy and the possibility to apply numerical methods to solve it. \change{We stress that, from the numerical viewpoint, well-posedness is highly desirable, since it entails that small errors in the initial data imply controllable errors in the numerical solution.}\footnote{Note that the requirement of well-posedness also applies to the case of ordinary differential equations (ODEs). %, but in that case it is enough to have smooth, i.e., $C^\infty$, coefficients to guarantee it \cite[p.~89]{Friedrich:2000qv}. 
In that case, the well-posedness of the Cauchy problem is established by the Picard--Lindel\"{o}f theorem \cite{picard,lindelof,lindelof2} (also called the Picard--Lipshitz theorem, or the fundamental theorem of ODEs), see also \cite[Chapter.~1]{grant2014theory}.}

It is important to emphasize that, contrary to Hadamard's opinion, ill-posed systems can in fact be physically relevant (see e.g. \cite{tikhonov1977solutions}). %For example, they arise in hydrodynamics \cite[p.~230]{courant1962methods} and in climate physics \cite{vonWurtemberg425588}. 
Examples of ill-posed problems can be found in \cite{Kabanikhin:2008}. However, in these cases the ill-posedness is acceptable, as it has to do with the physical description of the system. This is not the case in relativity (general or bimetric).

Following \cite[Sec.~11.1]{baumgarte2010numerical}, we now give the definition of a strongly hyperbolic system containing first-order time and spatial derivatives. The generalization to systems involving second-order spatial derivatives---as the (c)BSSN system in GR and BR---can be found in \cite{Gundlach_2006}. Consider the system of PDEs given by,
\begin{align}
\label{eq:PDE}
	\dt u^i + A^{k}{}^i{}_j\partial_k u^j = S^i,
\end{align}
where $u^i$ is the $n$-dimensional vector containing the unknown functions to solve for, $S^i$ is the source $n$-vector and each of the $A^{k}$ are $n\times n$ matrices of constant coefficients. Consider an arbitrary unit covector $n_{k}$, and define the \qm{principal symbol} or \qm{characteristic matrix} of the system \eqref{eq:PDE} as,
\begin{align}
\label{eq:principal}
	P^i{}_j \coloneqq n_{k} A^{k}{}^i{}_j.
\end{align}
The system \eqref{eq:PDE} is called \qm{strongly hyperbolic} if, for all unit covectors $n_{k}$, the principal symbol $P^i{}_j$ has real eigenvalues and a complete set of eigenvectors.  Strongly hyperbolic \change{first-order} systems \change{of PDEs lead to} well-posed \change{Cauchy problems} \cite[Theorem~6.2.2]{Kreiss1989ivp}.

\section{The bimetric field equations and their standard \nPlusOne formulation}
\label{sec:3p1}

The BFE can be written \cite{Hassan_2013},\footnote{We choose the same sign convention as in \cite{Kocic:2018ddp}, for the sign in front of the bimetric interaction potential in the action.}
\begin{align}
\label{eq:BFE}
	\gEinst^\mu{}_\nu 				&= \gKappa \rb{\gVse^\mu{}_\nu + \gTse^\mu{}_\nu}, \qquad
	\fEinst^\mu{}_\nu	=\fKappa \rb{\fVse^\mu{}_\nu +  \fTse^\mu{}_\nu},
\end{align}
where $\gEinst$ and $\fEinst$ are the Einstein tensors for the two metrics $\gMet$ and $\fMet$, $\gTse$ and $\fTse$ are two independent stress--energy tensors, $\gVse$ and $\fVse$ are the bimetric stress--energy tensors that couple the metrics, and $\gKappa$ and $\fKappa$ are the two Einstein gravitational constants. 
The bimetric stress--energy tensors may or may not satisfy the energy conditions (for the energy conditions, see \cite{hawking1973large}). In particular, note that in vacuum, if one of them satisfies the null energy condition, the other does not \cite{Baccetti2012}. When one of them does, it can be interpreted as a stress--energy tensor arising from the other spin-2 field. One example of this can be found in \cite{Kocic:2017hve}.

A fundamental feature about the bimetric stress--energy tensors is that they are non-derivative, i.e., they are not defined in terms of the derivatives of the metrics, but are functions of the square root matrix $\mSqrt=\rb{\gMet^{-1}\fMet}^{1/2}$ only, where the principal branch of the matrix square root function is assumed \cite{HassanKocic2018}. This allows us to rewrite the BFE \eqref{eq:BFE} as,
\begin{align}
	\gEinst^\mu{}_\nu 					&= \gTeff{}^\mu{}_\nu \qquad
	\fEinst^\mu{}_\nu	 	= \fTeff{}^\mu{}_\nu,
\end{align}
with $\gTeff{}^\mu{}_\nu\coloneqq\gVse^\mu{}_\nu + \gTse^\mu{}_\nu$ and $\fTeff{}^\mu{}_\nu\coloneqq\fVse^\mu{}_\nu +  \fTse^\mu{}_\nu$.

The BFE are then formally equivalent to two sets of Einstein equations coupled via the effective stress--energy tensors only. Then, one can recast them in the standard \nPlusOne decomposition by following the same steps as in GR. One defines a spacelike hypersurface embedded in the spacetime, where the initial data are specified, and projects the Einstein equations both on the hypersurface and on the direction orthogonal to it (see, e.g., \cite{baumgarte2010numerical,gourgoulhon20123+1}). The sources are now the sum of the external matter sources and the bimetric stress--energy tensors, whose decomposition has to be computed. This was done independently in \cite{Hassan:2011zd,HassanLundkvist2018} and \cite{Kocic:2018ddp} following different approaches. The result is a set of evolution equations, a set of constraints similar to those of GR, and the bimetric conservation law (BCL) $\mS=0$ (the so-called secondary constraint), which is crucial in eliminating the Boulware--Deser ghost \cite{PhysRevD.6.3368}, as explained in \cite{HassanLundkvist2018}. Here, we follow the approach in \cite{Kocic:2018ddp}, to which we refer the reader for the details. The \nPlusOne BFE computed in \cite{Kocic:2018ddp} are written explicitly in Appendix~\ref{app:stdeqs}.

The bimetric conservation law $\mS=0$ must be preserved in time, therefore $\partial_t \mS=0$. This is called the \qm{preservation of the bimetric conservation law}, and provides a relation between the lapse functions of the two metrics, of the form \cite{Alexandrov2013,HassanLundkvist2018,Kocic_2019}
\begin{equation}
\label{eq:ratio}
	\dfrac{\gLapse}{\fLapse}=-\dfrac{\sfW}{\sgW}\eqqcolon\mW =\mbox{scalar field independent of the lapses and the shifts}.
\end{equation}
Unfortunately, the explicit expression for $\mW$ is very complicated, even in spherical symmetry. See \cite{Kocic_2019} and the addendum to this paper in the ancillary files for its expression in the standard \threePlusOne and cBSSN formalisms, respectively. Note that the existence of the relation between the lapses is consistent with the fact that bimetric relativity is invariant under the action of a diffeomorphism group acting in the same way on both sectors. In other words, we are free to choose \emph{one} coordinate system for both metrics. In the \nPlusOne decomposition, this translates in the freedom to choose \emph{one} lapse function and \emph{one} shift vector only, as explained in more detail in \autoref{sec:gauge}.

\section{The covariant BSSN formulation of the bimetric field equations}
\label{sec:BSSN}

This section is devoted to the discussion of the BSSN and cBSSN formulation of the standard \threePlusOne BFE. We will mainly focus on the differences compared with GR. The explicit equations are presented in Appendix~\ref{app:BSSNbimint}, Appendix~\ref{app:BSSNeqs} and Appendix~\ref{app:cBSSNeqs}.

\subsection{The BSSN formulation}
\label{subsec:BSSN}

When rewriting the bimetric \threePlusOne equations [eqs.~\mref{eq:BEE,eq:BCE,eq:SC}] in the BSSN formulation, the starting point is the definition of the conformal metrics \cite[Sec.~11.5]{baumgarte2010numerical}
\begin{subequations}
\label{eq:confmetrics}
	\begin{alignat}{3}
		\gSpBS_{ij}	&\coloneqq \ee^{-4\gconf}\gSp_{ij},\qquad  \gSpBS^{ij} \  &\coloneqq \ee^{4\gconf}\gSp^{ij},	\\
		\fSpBS_{ij}		&\coloneqq \ee^{-4\fconf}\fSp_{ij},\qquad  \fSpBS^{ij}		&\coloneqq \ee^{4\fconf}\fSp^{ij},
	\end{alignat}
\end{subequations}
which are assumed to have determinants equal to 1. This renders the conformal metrics tensor densities of weight $-2/3$, and the conformal factors scalar densities of weight $1/6$. The conformal structure implies
	\begin{align}
	\label{eq:conffactors}
		\gconf = \dfrac{1}{12}\log (\gdet), 	\qquad	\fconf = \dfrac{1}{12}\log (\fdet).
	\end{align}
We also decompose the traceless part of the extrinsic curvatures as follows
\begin{subequations}
\label{eq:confA}
\begin{align}
	\gABS_{ij}	&\coloneqq\ee^{-4\gconf}\gEA_{ij}=\ee^{-4\gconf}\rb{\gK_{ij}-\dfrac{1}{3}\gSp_{ij}\gK}, \\
	\fABS_{ij}	&\coloneqq\ee^{-4\fconf}\fEA_{ij}=\ee^{-4\fconf}\rb{\fK_{ij}-\dfrac{1}{3}\fSp_{ij}\fK}.
\end{align}
\end{subequations}
Likewise, the conformal traceless extrinsic curvatures are tensor densitites of weight $-2/3$. The indices of the conformal tensors are raised and lowered by the conformal metrics of the corresponding metric sector. In the BSSN formulation, the \qm{conformal connections} are defined as,
\begin{align}
\label{eq:BGC}
	\gGBS^i \coloneqq \gSpBS^{jk}\gGBS^i_{jk}, \qquad \fGBS^i \coloneqq \fSpBS^{jk}\fGBS^i_{jk},
\end{align}
where $\gGBS^i_{jk},\fGBS^i_{jk}$ are the Christoffel symbols of the conformal metrics $\gSpBS_{ij},\fSpBS_{ij}$. The conformal connections transform as in eq.~(11.45) in \cite{baumgarte2010numerical}. The new dynamical variables in the BSSN formulation are the conformal metrics $\gSpBS_{ij}, \fSpBS_{ij}$, the conformal factors $\gconf,\fconf$, the traces of the extrinsic curvatures $\gK,\fK$, the conformal traceless parts of the extrinsic curvatures $\gABS_{ij},\fABS_{\ij}$ and the conformal connections $\gGBS^i,\fGBS^i$. With the appropriate transformation rules for these geometrical objects, the BSSN equations are covariant under spatial coordinate transformations not involving the time coordinate \cite{PhysRevD.79.104029}.

In order to evolve the system in time, one needs to choose a gauge. The BSSN equations are strongly hyperbolic if one chooses the standard gauge, introduced below, and enforce that $\gABS_{ij}$ is traceless during the evolution \cite{PhysRevD.66.064002,PhysRevD.70.104004,PhysRevD.79.104029}. The standard gauge consists in the \onePlusLog slicing for the lapse $\gLapse$ \cite{PhysRevLett.75.600}, and the $\Gamma$-driver condition for the shift $\gShift$ \cite{PhysRevD.67.084023},
\begin{subequations}
\label{eq:stdgauge}
	\begin{align}
		\dt\gLapse		&=\gShift^j\partial_j\gLapse-2\gLapse\gK, \\
		\dt\gShift^i	&=\gShift^j\partial_j\gShift^i+\dfrac{3}{4}\gShiftB^i \label{eq:Gamma1}\\
		\dt\gShiftB^i	&=\gShift^j\partial_j\gShiftB^i+\dt\gGBS^i-\gShift^j\partial_j\gGBS^i-\eta\gShiftB^i,\label{eq:Gamma2}
	\end{align}
\end{subequations}
where $\gShiftB^i$ is an auxiliary variable and $\eta$ is a freely specifiable real constant. As explained in \cite{PhysRevD.79.104029}, the $\Gamma$-driver condition \mref{eq:Gamma1,eq:Gamma2} is not spatially covariant. Suppose we have some initial data on the spacelike hypersurface, written in Cartesian coordinates. We can rewrite them in spherical coordinates by using the transformation rules for the tensors densities and the conformal connections. We then evolve these initial data according to the BSSN equations with the standard gauge, up to some time $t_f$. Since the $\Gamma$-driver condition is not covariant, the dynamical variables at $t_f$ are \emph{not} related by the same transformation rules for tensor densities and conformal connections. Therefore, the dynamical variables at $t_f$ in Cartesian coordinates and spherical coordinates differ geometrically. This problem can be solved by rewriting the BSSN equations and the standard gauge according to a procedure presented in \cite{PhysRevD.79.104029} (see also \cite{Alcubierre2011}) and summarized in \autoref{subsec:cBSSN}.

In BR, the rewriting of the evolution equations (BEE) and the constraint equations (BCE) in terms of the BSSN dynamical fields mimics the analog computation in GR, whereas the rewriting of the BCL has no analog in GR. Actually, the parts of the equations not involving the bimetric interactions are exactly the same as in GR. The bimetric BSSN constraint and evolution equations are listed in Appendix \ref{app:BSSNeqs}. The differences compared with GR are:
\begin{enumerate}
	\item The presence of the bimetric conservation law $\mS=0$.
	\item The fact that the lapse function and the shift vector of one of the metrics $\gMet$ and $\fMet$ are not freely specifiable.
	\item The effective sources include both the contribution from the external matter sources and the bimetric sources.
\end{enumerate}
The parts involving the bimetric interactions can be rewritten in the BSSN formulation by determining how the spatial vielbein $\gE,\fE$ of the spatial metrics $\gSp,\fSp$ transform under the conformal change \eqref{eq:confmetrics}. This is discussed in the next subsection.

\subsection{The BSSN formulation of the bimetric interactions}

The \nPlusOne decomposition of the BFE as formulated in \cite{Kocic:2018ddp} relies on the parametrization with respect to the geometric mean metric $\hMet=\gMet \operatorname{\#} \fMet\coloneqq\gMet\rb{\gMet^{-1}\fMet}^{1/2}$ of the metrics $\gMet$ and $\fMet$.\footnote{In index notation we have $\hMet_{ij} = \gMet_{ik}\qb{\rb{\gMet^{-1}\fMet}^{1/2}}{}^k{}_j$.} In this parametrization, the spatial metrics are written in terms of their vielbeins. In matrix notation,
	\begin{align}
		\gSp =\transpose{\gE} \sEta \gE, \qquad \fSp =\transpose{\fE} \sEta \fE,
	\end{align}
where the spatial vielbein $\gE$ is freely specifiable and the spatial vielbein $\fE$ is defined as
	\begin{alignat}{4}
	\label{eq:symcondition}
		\fE	&\coloneqq \sRs \fEtr, \quad & \quad \sEta^{-1}\sRs^\tr\sEta &=\sRs^{-1}, \nonumber \\
		 \sRs	&= \rb{\sEta^{-1}\transpose{\sRbar}\sEta\sRbar}^{1/2}\sRbar^{-1}, \quad & \sRbar &\coloneqq \sEta^{-1}\fEtr^{-1,\mathsmaller{\mathsf{T}}}\transpose{\gE}\sEta \sLs,
	\end{alignat}
where the freely specifiable vielbein $\fEtr$ of $\fSp$ is rotated into $\fE$ by the orthogonal transformation $\sRs$, and $\sEta$ is the spatial part of the Minkowski metric, i.e., the Euclidean metric.\footnote{For the sake of clarity, we write $\sEta$ explicitly in every equation where it appears, because it is needed to raise and lower the Lorentz indices.} The transformation $\sRs$ is determined by the requirement that the geometric mean $\hMet$ exists \cite{MikicaB.794295}. The operator $\sLs$ is the spatial part of a Lorentz boost with boost vector $\sLv =\sLt^{-1}\sLp$ and Lorentz factor $\sLt=(1+\transpose{\sLp}\sEta \sLp)^{1/2}$. It can be written as $\sLs=(\sI+\sLp\transpose{\sLp}\sEta)^{1/2}=\sI+\sLp\transpose{\sLp}\sEta/(\sLt+1)$, and the 4-dimensional Lorentz boost itself can be written as
\begin{align}
\label{eq:Lboost}
	\lSector{\boldsymbol{\Lambda}}=
	\begin{pmatrix}
		\sLt & \sLp^\tr\sEta \\
		\sLp & \sLs
	\end{pmatrix}.
\end{align}
See Appendix~\ref{app:BSSNbimint} for more details. In this framework, the real spatial vector $\sLp$, called \qm{separation parameter}, defines the shift vectors $\gShift$ and $\fShift$, respectively of the metrics $\gMet$ and $\fMet$, in terms of the shift vector $\hShift$ of the geometric mean metric $\hMet$,
	\begin{align}
	\label{eq:shifts}
		\gShift\coloneqq \hShift+\gLapse\, \sgn = \hShift+\gLapse\,\gE^{-1}\sLp\sLt^{-1}, \qquad \fShift\coloneqq \hShift-\fLapse\, \sfn = \hShift-\fLapse\,\fE^{-1}\sLp\sLt^{-1}.
	\end{align}
This is the most general parametrization of the bimetric \nPlusOne decomposition \cite{Kocic:2018ddp}.

Since all the spatial bimetric interactions terms, defined in \cite{Kocic:2018ddp} and reported in Appendix~\ref{app:BSSNbimint}, depend on the spatial vielbeins, on $\sLs$ and $\sRs$, their rewriting in the BSSN formalism relies on the conformal decomposition of these variables. The conformal decomposition of the vielbeins read,
\begin{align}
\label{eq:vielbeinBSSN}
	\gSpBS=\ee^{-4\gconf}\gSp \Longrightarrow \transpose{\gEBS}\, \sEta\, \gEBS=\rb{\ee^{-2\gconf}\transpose{\gE}} \sEta \rb{\ee^{-2\gconf}\gE},
\end{align}
which tells us that
\begin{subequations}
\label{eq:vielbeinBSSN2}
\begin{alignat}{2}
\gEBS	&=\ee^{-2\gconf}\gE, \qquad &\fEBS&=\ee^{-2\fconf}\fE, \\
\gEBS^{\,-1}&=\ee^{2\gconf}\gE^{-1}, \qquad &\fEBS^{-1}&=\ee^{2\fconf}\fE^{-1}.
\end{alignat}
\end{subequations}
From \eqref{eq:vielbeinBSSN2} and the conformal decomposition of $\sLs$ and $\sRs$, reported in Appendix~\ref{app:BSSNbimint}, we can derive the BSSN formulation of all the spatial bimetric interactions and sources. The explicit derivations are presented in Appendix~\ref{app:BSSNbimint}.

\subsection{The covariant extension of the BSSN formulation}
\label{subsec:cBSSN}

As we outlined in \autoref{subsec:BSSN}, the BSSN formulation with the standard gauge is not spatially covariant. As described in \cite{PhysRevD.79.104029}, this is a problem when comparing the same physical system in different coordinates. Therein, the BSSN formulation was generalized making it spatially covariant, obtaining the cBSSN formulation. Since the computations in \cite{PhysRevD.79.104029} do not alter the expressions of the matter sources in the evolution equations, the covariant generalization applies to both metric sectors in bimetric relativity.\footnote{The computations in \cite{PhysRevD.79.104029} are made in vacuum, but it is straightforward to add an external source and generalize them.} As a consequence, the bimetric sources have the same expression as in the BSSN formulation, given by \mref{eq:biminteractionsBSSN,eq:biminteractionsBSSN2,eq:bimsourcesBSSN}.\footnote{As a general rule in this context, every computation that does not concern the matter sources in the evolution and constraint equations in GR can be directly translated in BR without modification.}

Having a covariant version of the BSSN formulation is important in BR, since it allows us to safely use spherical coordinates. Since the Birkhoff theorem is not valid in BR, see, e.g, \cite{Kocic:2017hve}, a spherically symmetric solution of the BFE does not need to be static. From one side, this may be interpreted as an undesired feature of the theory; on the other hand, it makes the study of spherically symmetric systems much more interesting in BR than in GR, as discussed in the Introduction. Specifically, spherically symmetric systems in BR emit longitudinally polarized gravitational radiation, which can be tested against observational data.

In addition, using spherical coordinates made it possible to compute both the ratio between the lapses $\mW$ appearing in \eqref{eq:ratio} and the spatial Lorentz vector $\sLp$ appearing in \eqref{eq:shifts} in the difference between the shifts (see \cite{Kocic:2018ddp,Kocic_2019} and Appendix \ref{app:sphsym} for more details). Note that $\sLp$ is also known in the most general $\beta_{(1)}$-model \cite{Hassan:2011tf,HassanLundkvist2018}, where the explicit expression for $\sfn=\fE^{-1}\sLp\sLt^{-1}$ in \eqref{eq:shifts} is computed.\footnote{Knowing $\sfn$, $\sLp=\big[1-(\fE\sfn)^\tr(\fE\sfn)\big]^{-1/2}\fE\sfn$.}

In the BSSN formulation, the determinant of the conformal metric $\gSpBS$ is taken to be $1$, making it a scalar rather than a scalar density. This also alters the transformation properties of the metric and the extrinsic curvature, making them tensor densities. In the cBSSN formulation, nothing is assumed on the transformation properties of the determinant of the conformal metric. The new conformal decomposition of the metrics and the extrinsic curvatures becomes,
\begin{subequations}
\label{eq:confmetricscov}
	\begin{alignat}{3}
		\gSpBS_{ij}	&\coloneqq \ee^{-4\gconf}\gSp_{ij},\qquad  \gSpBS^{ij} \  &\coloneqq \ee^{4\gconf}\gSp^{ij},	\\
		\fSpBS_{ij}		&\coloneqq \ee^{-4\fconf}\fSp_{ij},\qquad  \fSpBS^{ij}		&\coloneqq \ee^{4\fconf}\fSp^{ij},
	\end{alignat}
\end{subequations}
which is the same as before, without the restriction on the determinants $\gdetBS$ and $\fdetBS$, and
\begin{subequations}
\label{eq:confAcov}
\begin{align}
	\gABS_{ij}	&\coloneqq\ee^{-4\gconf}\gEA_{ij}=\ee^{-4\gconf}\rb{\gK_{ij}-\dfrac{1}{3}\gSp_{ij}\gK+\dfrac{1}{3}\gSp_{ij}\gABS}, \\
	\fABS_{ij}	&\coloneqq\ee^{-4\fconf}\fEA_{ij}=\ee^{-4\fconf}\rb{\fK_{ij}-\dfrac{1}{3}\fSp_{ij}\fK+\dfrac{1}{3}\fSp_{ij}\fABS}, \\
	\gKBS		&\coloneqq \gK-\gABS , \qquad    \fKBS \coloneqq \fK-\fABS.
\end{align}
\end{subequations}
The tensors $\gABS$ and $\fABS$ are not traceless, as can be seen by comparing \eqref{eq:confAcov} with \eqref{eq:confA}. The conformal connections in \eqref{eq:BGC} are made covariant by introducing two background connections $\gGbackBS^i_{jk},\fGbackBS^i_{jk}$ \cite{Garfinkle_2008}. It is possible, but not necessary, to introduce two background metrics whose Levi--Civita connections serve as the background connections.
We define the new dynamical variables,
\begin{align}
\label{eq:confconcov}
	\gL^i \coloneqq \gSp^{jk}\gDG^i_{jk}=\gSp^{jk}\Bigl(\gGBS^i_{jk}-\gGbackBS^i_{jk}\Bigr), \qquad \fL^i \coloneqq \fSp^{jk}\fDG^i_{jk}=\fSp^{jk}\Bigl(\fGBS^i_{jk}-\fGbackBS^i_{jk}\Bigr).
\end{align}
Since the difference between two Christoffel symbols is a tensor, our set of dynamical variables includes tensors only, i.e., $\gSp_{ij},\fSp_{ij},\gconf,\fconf,\gABS_{ij},\fABS_{ij},\gKBS,\fKBS,\gL,\fL$. The standard gauge \eqref{eq:stdgauge} in the cBSSN formalism reads,
\begin{subequations}
\label{eq:standardgauge}
	\begin{align}
		\label{eq:1plog}\partial_t \gLapse 		&= \gShift^j \gDback_j\gLapse -2\gLapse \gKBS, \\
		\label{eq:gdriver1}\partial_t \gShift^i	 	&= \gShift^j \gDback_j\gShift^i +\dfrac{3}{4}\gShiftB^i, \\
		\label{eq:gdriver2}\partial_t \gShiftB^i 	&= \gShift^j \gDback_j\gShiftB^i +\rb{\partial_t\gL^i -\gShift^i \gDback_j\gL^i} -\eta \gShiftB^i,
	\end{align}
\end{subequations}
which is manifestly covariant. The explicit bimetric cBSSN constraint and evolution equations are written in Appendix \ref{app:cBSSNeqs}.

We emphasize that the background connections are completely arbitrary, and in GR, there is no preferred connection to be chosen. In bimetric relativity, a third metric $\hMet$ is defined.\footnote{Actually, in the space of pseudo-Riemannian metrics built on our manifold, we have a path of metrics connecting $\gMet$ and $\fMet$, corresponding to a geodesic of the trace metric and parameterized by $h_\alpha = \gMet\rb{\gMet^{-1}\fMet}^\alpha, \;\alpha \in \mathbb{R}$ (see \cite{PhysRevD.97.084022,Kocic:2018ddp}).} Hence, we can choose the Levi--Civita connection of the conformal spatial metric $\hSpBS$ to define the covariant conformal connection in \eqref{eq:confconcov}. The consequences of this choice are described in more detail in \cite{Torsello_2019}.

In the cBSSN formulation, the determinants $\gdetBS,\fdetBS$ of the conformal metrics and the traces $\gABS,\fABS$ are left undetermined, making the cBSSN evolution equations in \mref{eq:cBSSNEEg, eq:cBSSNEEf} incomplete \cite{PhysRevD.79.104029}. We must choose how these quantities evolve, in order to be able to evolve the full system. Following \cite{PhysRevD.79.104029}, we choose $\dt \gABS = \dt \fABS=0$. Regarding the determinants, there are two natural choices, referred to as \qm{Lagrangian} and \qm{Eulerian} \cite{PhysRevD.71.104011,Brown_2008,PhysRevD.79.104029}, given by
\begin{align}
	\dt \gdetBS =0, \qquad \pfg \gdetBS =\dt \gdetBS-\mathscr{L}_\gShift \gdetBS = \dt \gdetBS -2\gdetBS\gDBS_i\gShift^i =0,
\end{align}
with analog expressions for the $\fMet$-sector. In BR, we need to specify the evolution of the determinants and the traces in both sectors. More details on this can be found in \cite{Torsello_2019}. The expression for $\dt\gL^i$ in \eqref{eq:gdriver2} is explicitly substituted with the Lagrangian formulation of \eqref{eq:cBSSNEEgL}.

\section{Well-posedness and gauge choices}
\label{sec:gauge}

As established in \cite{Kocic:2018ddp}, the three lapses and shifts of $\gMet,\fMet$ and $\hMet$ are related by,
\begin{subequations}
\label{eq:gaugevariables}
	\begin{alignat}{3}
		\gLapse^2 &=\hLapse ^2 \sLt\mW, \qquad & \fLapse ^2 &= \dfrac{\hLapse ^2 \sLt}{\mW}, \qquad & \gLapse &=\mW \fLapse ,\label{eq:gaugevariableslapses} \\
		\gShift &=\hShift + \dfrac{\gLapse}{\sLt} \gE^{-1}\sLp , \qquad & \fShift &=\hShift - \dfrac{\fLapse}{\sLt} \fE^{-1}\sLp, \qquad & \fShift &=\gShift -\dfrac{\fLapse}{\sLt}\rb{ \mW\,\gE^{-1}+\fE^{-1} }\sLp,
	\end{alignat}
\end{subequations}
where $\sLt = \sqrt{1+\sLp^\tr \sEta \sLp}$ and $\hLapse$ is the lapse of $\hMet$.
We are free to choose one lapse function and one shift vector, exactly as in GR; the other four quantities are determined by \eqref{eq:gaugevariables}.

When imposing a gauge in BR, it can be written in terms of any of the three lapse functions or the shift vectors. Suppose, for example, that we choose the standard gauge with respect to $\gMet$ in \eqref{eq:standardgauge}. This gauge can be rewritten in terms of the lapse functions $\fLapse,\hLapse$ and the shift vectors $\fShift,\hShift$ by using \eqref{eq:gaugevariables}. Hence, we can impose the standard gauge with respect to $\gMet$ by gauge fixing, say, $\hLapse$ and $\fShift$. We say that we \qm{choose a gauge condition with respect to a metric}, to emphasize that the geometry of the slicing is determined by that metric. In addition, we say that we \qm{gauge fix} one of the lapses and shifts. The same gauge choice (or gauge condition) can be expressed via different, but equivalent, gauge fixings. It follows that, in BR one can have \qm{mixed} gauges, i.e., one can choose the \onePlusLog slicing with respect to $\hSp$, and the $\Gamma$-driver condition with respect to $\fSp$. In this case, $\hMet$ would determine the time slicing, whereas $\fMet$ would determine the spatial gauge. If these gauges are singularity avoiding or horizon penetrating for any of the metrics remains an open question. See \cite{Torsello_2019} for a study of the \qm{mean gauges}, i.e., the gauge choices with respect to the mean metric $\hMet$.

In GR, the cBSSN formulation is strongly hyperbolic if one chooses the standard gauge and fulfills some other technical conditions \cite{PhysRevD.79.104029}. In BR, the well-posedness of the evolution equations involves both of the metric sectors. Suppose that we fix the lapse and shift of one metric to be determined by the standard gauge. Now, the bimetric source $\gJotab^i{}_j$ appearing in the evolution equation for the conformal extrinsic curvature \eqref{eq:bimexplicitsources} contains the ratio of the lapses $\mW$. The general explicit expression of $\mW $ is not known, but it can be computed explicitly in spherical symmetry (\cite{Kocic_2019} and the addendum to this paper). In that case, $\mW $ is a lengthy expression---roughly, it fills two pages---which depends on the radial derivatives of the dynamical fields. If the radial derivatives can not be replaced by algebraic expressions, $\mW$ will affect the characteristic matrix in \eqref{eq:principal} and alter the hyperbolic structure of the equations in the $\gMet$-sector compared with GR. Following the procedure described in \cite{ALCUBIERRE200576}, which promotes the logarithmic radial derivatives of the dynamical fields to be new dynamical variables\change{, thus achieving a first-order reduction of the system}, one ends up with an expression for $\mW$ which only depends on three radial derivatives, namely,
\begin{align}
	\dr \po, \quad \dr \bigg(\gAt+\dfrac{1}{3}\gKBS\bigg), \quad \dr \bigg(\fAt+\dfrac{1}{3}\fKBS\bigg).
\end{align}
Here, $\po$ is the only nonzero component of $\sLp$ and $\gAt,\fAt$ are the $\gABS^\theta{}_\theta,\fABS^\theta{}_\theta$ components of the conformal extrinsic curvatures (see Appendix \ref{app:sphsym} for more details). By using the two momentum constraints \mref{eq:cBSSNBCEsphMCg,eq:cBSSNBCEsphMCf} and the bimetric conservation law \eqref{eq:bimconssphsym} rewritten according to the procedure in \cite{ALCUBIERRE200576}, these three derivatives can be substituted with expressions involving the dynamical fields only algebraically. In more detail, the BCL can be solved for $\dr\po$, and the momentum constraints can be solved for $\dr(\gAt+\gKBS /3)$ and $\dr(\fAt+\fKBS /3)$.\footnote{Note that, in this way, we can freely specify $\sLp$ on the initial hypersurface, and the value of these three derivatives will depend on this choice.} This means that the ratio of the lapses $\mW$ is a purely algebraic expression in the dynamical fields, and, as such, does not enter the characteristic matrix in \eqref{eq:principal} and does not alter the hyperbolicity of the equations in the $\gMet$-sector compared with GR. Therefore, if we choose the standard gauge \eqref{eq:standardgauge} for $\gLapse$ and $\gShift$, the equations in the $\gMet$-sector are strongly hyperbolic.

Note that the algebraicity of $\mW$ in the dynamical fields is compatible with the fact that, in the case of static spherically symmetric black hole solutions in BR (see \cite{PhysRevD.85.124043,Brito:2013xaa,PhysRevD.96.064003} and reference therein) the function describing the ratio of the lapses, $\tau$ in \cite{PhysRevD.96.064003}, is determined by an algebraic relation. This relation corresponds precisely to the PBCL $\dt\mS=0$. Also, this confirms that the bimetric stress--energy tensors (of which the spatial bimetric interactions are the projections \cite{Kocic:2018ddp}) are non-derivative and cannot spoil the hyperbolic structure of the equations compared with GR (in spherical symmetry).

On the other hand, we need to consider the $\fMet$-sector as well. The equations in the $\fMet$-sector formally appear the same as in the $\gMet$-sector, but now the lapse and shift of $\fMet$ are determined by \eqref{eq:gaugevariables}. As a consequence, the ratio of the lapses appears wherever the lapse and shift of $\fMet$ appear. Since there are terms involving first and second spatial derivatives of the lapse and shift, they contain first and second spatial derivatives of $\mW$. Hence, they contain the spatial derivatives of \emph{all} the dynamical fields, and we cannot eliminate all of them by using the constraints. Therefore, the hyperbolic structure of the PDEs in the $\fMet$-sector is drastically different compared with GR. \change{This means that one cannot carry over the GR results to BR, and additional steps are needed towards a definite answer regarding the strong hyperbolicity of the equations}.

Instead of using \eqref{eq:gaugevariables} to replace the lapse of $\fMet$, we could equivalently use it to determine its value on the initial hypersurface, and impose the preservation in time of the PBCL \eqref{eq:ratio} by setting its time derivative to 0. This gives an evolution equation for the lapse of $\fMet$, which becomes a dynamical variable.\footnote{More in general, this is an evolution equation for the lapse that we do not gauge fix.} Hence, with this choice, the principal symbol of the system of PDEs is largely different than in GR. For example, the evolution equation for the lapse of $\fMet$ involves the time derivative of the ratio of the lapses $\mW$ which nontrivially alters the hyperbolic structure compared with GR. See \cite[Sec.~6.2, p.~103]{Torsello1390334} for more details.

The analysis above holds in the case of spherical symmetry, which is presented in Appendix \ref{app:sphsym}. This study suggests that the ratio of the lapses, $\mW $, always contains some spatial derivatives of the dynamical fields which can be eliminated by using the constraints, but the hyperbolicity is altered compared to GR by the spatial derivatives of the lapses and shifts involving $\mW$. Therefore, the computation of $\mW$ in the general case is a prerequisite for the study of the well-posedness of any formulation of the BFE.

A possible gauge choice, which preserves the symmetry of the equations between $\gMet $ and $\fMet$ and modifies the hyperbolic structures of both sectors in a more symmetric---and hopefully better-behaved---way, is to fix the lapse and the shift of $\hMet$. In this case, we see from \eqref{eq:gaugevariables} that $\mW$ appears in the spatial derivatives of the lapses and shifts of both $\gMet$ and $\fMet$, thus modifying the hyperbolic structure in both sectors. This is investigated in more detail in \cite{Torsello_2019}, where both the standard gauge and the maximal slicing for $\hLapse$ and $\hShift$ are computed.

\section{Summary and outlook}
\label{sec:sum}

In this paper, we presented the covariant BSSN formalism of the bimetric field equations. We emphasized why this formulation is important in bimetric relativity and we stressed the differences with the analogous formulation of the Einstein equations, summarized below.
\begin{enumerate}
	\item In addition to the Hamiltonian and momentum constraints for both metrics, there is an additional bimetric conservation law (the so-called secondary constraint) $\mS=0$ that, in the free evolution scheme, has to be solved for the initial data on the spacelike hypersurface.
	\item The sources in the equations include both the external matter sources and the bimetric sources in \eqref{eq:bimexplicitsources}. After a first-order reduction of the PDEs, the bimetric sources do not contain the derivatives of the dynamical fields. Hence, they do not alter the hyperbolic structure of the equations compared to GR.
	\item Bimetric relativity is diffeomorphism invariant. This provides us with the possibility to choose \emph{one} lapse function and \emph{one} shift vector of any of the metrics, $\gSp$, $\fSp$ or their geometric mean $\hSp$. The remaining lapses and shifts are determined by \eqref{eq:gaugevariables}.
	\item The relation between the lapses is established in \cite{Alexandrov2013,HassanLundkvist2018} by imposing the preservation of the bimetric conservation law in time, $\dt \mS=0$, and it is computed explicitly in \cite{Kocic_2019} for spherically symmetric spacetimes in the standard \threePlusOne formulation. The expression in the covariant BSSN formalism is given in the addendum to this paper. The ratio between the lapses, $\mW$, is a lengthy algebraic expression in terms of the dynamical fields and their spatial derivatives. Since the evolution equations involve the spatial derivatives of $\mW$, the hyperbolic structure of the system of PDE is different compared with the corresponding equations in GR. The hyperbolic structure is changed in either one of the two metric sectors, or in both, depending on the gauge choice.
\end{enumerate}
Other than these four differences, the system is analogous to the covariant BSSN formulation of the Einstein field equations presented in \cite{PhysRevD.79.104029}. In particular, from the viewpoint of numerical relativity, the bimetric fields equations can be tackled numerically in the same way as the Einstein field equations. However, since \change{we showed that the results in GR cannot be carried over to BR in a straightforward way, and} the well-posedness of the problem is not proved yet, we do not know how successful this can be. Nonetheless, the equations do offer some stimulating challenges:
\begin{enumerate}
	\item The computation of $\mW$ and $\sLp$ is necessary to be able to solve the bimetric equations in any formulation. At present, $\mW$ is only computed under the assumption of spherical symmetry, whereas $\sLp$ is computed in spherical symmetry and in the most general $\beta_{(1)}$-model; we refer the reader to \cite{Kocic_2019,Hassan:2011tf,HassanLundkvist2018} for more details.
	\item Investigating if the bimetric covariant BSSN evolution equations, together with a suitable gauge, are strongly hyperbolic is of great importance and depends on the computation of $\mW$. Since the latter is known in spherical symmetry, one can study the hyperbolicity of the evolution equations in \eqref{eq:cBSSNEEsph}.
	\item The choice of the gauge is essential in bimetric relativity (as it is in GR as well). In \cite{Torsello_2019}, we study some possible gauge choices which alter the hyperbolic structure of the evolution equations in both sectors. In particular, we investigate the gauge fixing on the geometric mean metric $\hMet$.
	\item The challenge is to integrate the bimetric BSSN equations numerically in spherical symmetry, e.g., for a gravitational collapse of matter or a non-static black hole solution. The numerical computation of both $\mW$ and $\sLp$ significantly reduces the accuracy of the integration. We have written a Mathematica/C++ code to perform the simulations\change{, see \cite{Kocic:2019gxl} for results obtained using the standard \threePlusOne equations.} The \change{results concerning the cBSSN formulation} will be the subject of another work. We remind the reader that, since the Birkhoff theorem is not valid in bimetric relativity (see, e.g., \cite{Kocic:2017hve}), the spherically symmetric case is very interesting to study. One can look for both vacuum and non-vacuum spherical solutions with nontrivial dynamics, and perhaps gravitational wave emission. This can potentially lead to results that could directly be compared against observational data.
\end{enumerate}

\subsection*{Acknowledgments}

It is a pleasure to thank Anders Lundkvist and Fawad Hassan for many fruitful discussions. We thank Anders Lundkvist and Giovanni Camelio for providing valuable remarks after a careful reading of the paper.

\appendix

\section{Explicit equations and computations}
\label{app:abeqs}

\subsection{The BSSN decomposition of the bimetric interactions and sources}
\label{app:BSSNbimint}

Consider the spatial parts $\gSp=\gE^\tr\sEta\gE$ and $\fSp=\fEtr^\tr\sEta\fEtr$ of two Lorentzian metrics $\gMet,\fMet$. In \cite{MikicaB.794295}, it is established that the existence of the \emph{real} square root $\rb{\gMet^{-1}\fMet}^{1/2}$ implies,
\begin{subequations}
\label{eq:symcondcomplete}
	\begin{align}
		\gShift	&\coloneqq \hShift+\gLapse \sgn = \hShift+\gLapse\gE^{-1}\sLp\sLt^{-1},\label{eq:symcondgShi} \\
		\fShift	&\coloneqq \hShift-\fLapse \sfn = \hShift-\fLapse\fE^{-1}\sLp\sLt^{-1}, \label{eq:symcondfShi} \\
		\hSp 		&=\gE^\tr\sEta\sLs\sRs\fEtr = \hSp^\tr. \label{eq:symcondchi}
	\end{align}
\end{subequations}
To be more precise, the freely specifiable spatial vielbein $\fEtr$ is used to compute the vielbein %$\sLs
$\sRs\fEtr$ such that the spatial part $\hSp$ of the geometric mean metric $\hMet=\gMet\operatorname{\#}\fMet$ is given by $\hSp=\gE^\tr\sEta\sLs(\sRs\fEtr)$. This is obtained by imposing \eqref{eq:symcondchi} and solving it for the Euclidean orthogonal transformation $\sRs$ in terms of $\sLs$ and the vielbeins $\gE,\fEtr$. Such a solution always exists, as it is part of the polar decomposition of the invertible matrix $\sRbar$ \cite{MikicaB.794295,Hassan:2014gta} [see (3.7) ]. For the sake of simplicity, we define the new vielbein of $\fSp$ to be $\fE\coloneqq \sRs \fEtr$; % rather than $\sLs\sRs\fEtr$; 
we have the freedom to do that since %$\fSp=\fEtr^\tr\sRs^\tr\sLs^\tr\sEta\sLs\sRs\fEtr=
$\fEtr^\tr\sRs^\tr\sEta\sRs\fEtr=\fEtr^\tr\sEta\fEtr$, implying that $\fSp$ is blind to this choice. The interaction terms are not affected as well, since they always contain both $\sLs$ and $\sRs$, irrespective of this choice. The matrix $\sLs$ explicitly appears in them. On the contrary, $\sRs$ does not appear explicitly, but it is taken into account implicitly inside $\fE$.

We define the bimetric interactions as \cite{Kocic:2018ddp},
\begin{subequations}
\begin{alignat}{2}
\label{eq:stdbiminteractions}
	\sgn &\coloneqq \gE^{-1}\sLv, \quad & \sfn &\coloneqq \fE^{-1}\sLv, \\
	\sgQ &\coloneqq \gE^{-1}\sLs^2\gE, \quad & \sfQ &\coloneqq \fE^{-1}\sLs^2\fE, \\
	\sgD &\coloneqq \fE^{-1}\sLs^{-1}\gE, \quad & \sfD &\coloneqq \gE^{-1}\sLs^{-1}\fE, \\
	\sgB &\coloneqq \sgD^{-1}=\gE^{-1}\sLs\fE, \quad & \sfB &\coloneqq \sfD^{-1}=\fE^{-1}\sLs\gE, \\
	\sgV &\coloneqq -\betaSum e_n(\sfD), \quad & \sfV &\coloneqq -\sLt^{-1}\betaSum e_{n-1}(\sgB), \\
	\sgU &\coloneqq -\sLt^{-1}\betaSum Y_{n-1}(\sgB), \quad & \sfU &\coloneqq -\sfD\betaSum Y_{n-1}(\sfD), \\
	\sgQU &\coloneqq \sgQ\sfU= -\sgB\betaSum Y_{n-1}(\sfD), \quad & \sfQU &\coloneqq \sfQ\sgU= -\sLt^{-1}\sfQ\betaSum Y_{n-1}(\sgB),\label{eq:sQU}
\end{alignat}
\end{subequations}
where $e_n(X)$ are the elementary symmetric polynomials of the linear operator $X$,
\begin{align}
	e_n(X)=X^{[a_1}{}_{a_1} X^{a_2}{}_{a_2}\dots X^{a_n]}{}_{a_n},
\end{align}
and $Y_n(X)$ is defined as,
\begin{align}
	Y_n(X)\coloneqq\sum_{k=0}^n(-1)^{n+k}e_k(X)X^{n-k}.
\end{align}
See \cite{Kocic:2018ddp} for more details about the properties of $e_n(X)$ and $Y_n(X)$. Note that $\stdim$ is the dimension of the spacetime, i.e., $\stdim=$ \nPlusOne. Hence, some terms in the summations will be zero. The $\beta_{(n)}$ parameters are $d+1$ real dimensionless constants appearing in the bimetric interaction potential, together with the energy scale $m$ \cite{Hassan:2011zd}. We define the bimetric sources (respectively, the bimetric energy densitites, the bimetric currents and the bimetric spatial stress--energy tensors) as \cite{Kocic:2018ddp},
\begin{subequations}
\label{eq:bimexplicitsources}
	\begin{alignat}{3}
		\grhob		&= -e_n(\sgB),											& \quad \gjotab_i	&=-\gSp_{ik} \sgQU^k{}_j\sgn^j, 					&\quad \gJotab_{ij}		&= \gSp_{ik}\qb{\sgV \delta^k{}_j-\sgQU^k{}_j+\mW^{-1}\sgU^k{}_j}, \\
		\frhob		&= -\dfrac{\sLt e_{n-1}(\sgB)}{\mydet{\fE\gE^{-1}}},	& \quad \fjotab_i	&= -\dfrac{\gjotab_i}{\mydet{\fE\gE^{-1}}}, 	& \quad \fJotab_{ij}		&= \dfrac{\fSp_{ik}\qb{\sfV \delta^k{}_j-\sfQU^k{}_j+\mW\,\sfU^k{}_j}}{\mydet{\fE\gE^{-1}}},
	\end{alignat}
\end{subequations}
where the summation $-\betaSum$ is understood in front of all the bimetric sources. Note the relation between the two bimetric currents $\gjotab_i,\fjotab_i$, which implies the relation (A.35) between the momentum constraints .

Here we compute the expressions for the bimetric interaction and sources in the (c)BSSN formalism. We require that the symmetrization condition \eqref{eq:symcondcomplete} holds for the BSSN variables as well. Since the shifts are the same in the BSSN formalism, we require conditions \mref{eq:symcondgShi,eq:symcondfShi} to stay the same. The condition \eqref{eq:symcondchi} should instead lead to its analog in the BSSN formalism,
\begin{align}
	\hSpBS 		&=\gEBS^\tr\sEta\sLsBS\sRsBS\fEtrBS = \hSpBS^\tr,
\end{align}
where $\sLsBS,\sRsBS$ are the BSSN counterparts of the spatial part of the Lorentz boost (3.8) and the orthogonal transformation in (3.7) , whose expression is unknown yet.

We start by computing the conformal decomposition of the objects in the Lorentz frame. The requirement that \mref{eq:symcondgShi} stays the same implies,
\begin{align}
		\gShift	& = \hShift+\gLapse\gE^{-1}\sLp\sLt^{-1} = \hShift+\gLapse\gE^{-1}\xi\sLpBS\sLtBS^{-1} \iff \sLp\sLt^{-1}=\xi\sLpBS\sLtBS^{-1},
\end{align}		
where the scalar $\xi$ accounts for the conformal decomposition of $\sLp\sLt^{-1}$. It follows that,
\begin{align}
\label{eq:collp}
	\dfrac{\sLp}{(1+\sLp^\tr\sEta\sLp)^{1/2}}=\dfrac{\xi\sLpBS}{(1+\sLpBS^\tr\sEta\sLpBS)^{1/2}} \iff \sLp = \xi\rb{\dfrac{1+\sLp^\tr\sEta\sLp}{1+\sLpBS^\tr\sEta\sLpBS}}^{1/2}\sLpBS.
\end{align}
We apply $\sLp^\tr\sEta$ to \eqref{eq:collp} and obtain,
\begin{align}
\label{eq:norms}
	\dfrac{\sLp^\tr\sEta\sLp}{(1+\sLp^\tr\sEta\sLp)^{1/2}}=\dfrac{\xi\sLp^\tr\sEta\sLpBS}{(1+\sLpBS^\tr\sEta\sLpBS)^{1/2}}=\xi^2\rb{\dfrac{1+\sLp^\tr\sEta\sLp}{1+\sLpBS^\tr\sEta\sLpBS}}^{1/2}\dfrac{\sLpBS^\tr\sEta\sLpBS}{(1+\sLpBS^\tr\sEta\sLpBS)^{1/2}},
\end{align}
which is equivalent to
\begin{align}
\label{eq:normssimp}
	\dfrac{\sLp^\tr\sEta\sLp}{1+\sLp^\tr\sEta\sLp}=\dfrac{\xi^2\sLpBS^\tr\sEta\sLpBS}{1+\sLpBS^\tr\sEta\sLpBS} \iff \sLpBS^\tr\sEta\sLpBS=\dfrac{\sLp^\tr\sEta\sLp}{\xi^2(1+\sLp^\tr\sEta\sLp)-\sLp^\tr\sEta\sLp}.
\end{align}
Hence, in general, we can rescale $\sLp$ as in \eqref{eq:collp} with a generic $\xi$ when we recast the equations into the (c)BSSN formulation, as long as we satisfy \eqref{eq:normssimp}. However, there is no need to rescale it since this is an unnecessary complication. Indeed, we can always satisfy \eqref{eq:collp} and \eqref{eq:normssimp} by choosing $\xi=1$, which implies $\sLp=\sLpBS$. It immediately follows,
\begin{subequations}
\label{eq:boostBS}
\begin{align}
	\sLp		=\sLpBS		& \Longrightarrow \sLt=\rb{1+\sLp^\tr\sEta\sLp}^{1/2}=\rb{1+\sLpBS^\tr\sEta\sLpBS}^{1/2}=\sLtBS \\
								&\Longrightarrow \sLv = \sLp\sLt^{-1}= \sLpBS\sLtBS^{-1} = \sLvBS \\
								&\Longrightarrow \sLs = \rb{1+\sLp\sLp^\tr\sEta}^{1/2} = \rb{1+\sLpBS\sLpBS^\tr\sEta}^{1/2} = \sLsBS.\label{eq:sLsBS}
\end{align}
\end{subequations}
Let us now compute the BSSN version of $\sRbar$, introduced in \eqref{eq:symcondition}. From \mref{eq:vielbeinBSSN2,eq:sLsBS} and $\fEtrBS =\ee^{-2\fconf}\fEtr$ [which follows from \eqref{eq:vielbeinBSSN2}] it follows that,
\begin{align}
\label{eq:sRsBS1}
	\sRbar &\coloneqq \sEta^{-1}\fEtr^{-1,\tr}\gE^{\tr}\sEta\sLs= \ee^{2(\gconf-\fconf)}\sEta^{-1}\fEtrBS^{-1,\tr}\gEBS^{\tr}\sEta\sLs=\ee^{2(\gconf-\fconf)}\sRbarBS,
\end{align}
which implies,
\begin{align}
\label{eq:sRsBS2}
	\sRs &\coloneqq \rb{\sEta^{-1}\sRbar^\tr \sEta\sRbar}^{1/2}\sRbar^{-1} =\ee^{2\rb{\gconf-\fconf}}\rb{\sEta^{-1}\sRbarBS^\tr \sEta\sRbarBS}^{1/2}\ee^{-2\rb{\gconf-\fconf}}\sRbarBS^{-1}= \sRsBS.
\end{align}
%consistently with \eqref{eq:fvielbeins} and \eqref{eq:fconfvielbeins}. 
Using \mref{eq:vielbeinBSSN2,eq:sLsBS,eq:sRsBS2}, the spatial part of the symmetrization condition \eqref{eq:symcondchi} can be written as,
\begin{align}
\label{eq:symcondchiBS}
	\hSp 			&=\gE^\tr\sEta\sLs\sRs\fEtr = \ee^{2\rb{\gconf+\fconf}}\gEBS^\tr\sEta\sLs\sRs\fEtrBS\eqqcolon \ee^{2\rb{\gconf+\fconf}}\hSpBS \nonumber \\
					&=\hSp^\tr	= \rb{\gE^\tr\sEta\sLs\sRs\fEtr}^\tr = \ee^{2\rb{\gconf+\fconf}}\rb{\gEBS^\tr\sEta\sLs\sRs\fEtrBS}^\tr \eqqcolon \ee^{2\rb{\gconf+\fconf}}\hSpBS^\tr,
\end{align}
i.e., if $\hSp$ is symmetric, its BSSN counterpart $\hSpBS$ is also symmetric, as desired.

In light of \mref{eq:vielbeinBSSN2,eq:boostBS,eq:sRsBS2}, we compute the BSSN decomposition of bimetric interactions,
\begin{subequations}
\label{eq:biminteractionsBSSN}
\begin{alignat}{2}
	\sgn &\coloneqq \gE^{-1}\sLv = e^{-2\gconf}\gEBS^{-1}\sLv = e^{-2\gconf}\sgnBS, \qquad & \sfn &\coloneqq \fE^{-1}\sLv= e^{-2\fconf}\fEBS^{-1}\sLv = e^{-2\fconf}\sfnBS, \\
	\sgQ &\coloneqq \gE^{-1}\sLs^2\gE \quad & \sfQ &\coloneqq \fE^{-1}\sLs^2\fE \nonumber\\
			&=\ee^{-2\gconf}\rb{\gEBS^{-1}\sLs^2\gEBS}\ee^{2\gconf} & & =\ee^{-2\fconf}\rb{\fEBS^{-1}\sLs^2\fEBS}\ee^{2\fconf} \nonumber\\
			&=\gEBS^{-1}\sLs^2\gEBS=\sgQBS, & & =\fEBS^{-1}\sLs^2\fEBS=\sfQBS,\label{eq:sQBS} \\
	\sgD &\coloneqq \fE^{-1}\sLs^{-1}\gE \quad & \sfD &\coloneqq \gE^{-1}\sLs^{-1}\fE \nonumber\\
	&=\ee^{-2\fconf}\rb{\fEBS^{-1}\sLs^{-1}\gEBS}\ee^{2\gconf} & & =\ee^{-2\gconf}\rb{\gEBS^{-1}\sLs^{-1}\fEBS}\ee^{2\fconf} \nonumber\\
			&=\ee^{2\rb{\fconf-\gconf}}\rb{\gEBS^{-1}\sLs^{-1}\fEBS}=\ee^{2\rb{\gconf-\fconf}}\sgDBS, &\qquad & =\ee^{2\rb{\fconf-\gconf}}\rb{\gEBS^{-1}\sLs^{-1}\fEBS}=\ee^{2\rb{\fconf-\gconf}}\sfDBS, \\
	\sgB &\coloneqq \sgD^{-1}=\ee^{2\rb{\fconf-\gconf}}\sgBBS, \quad & \sfB &\coloneqq \sfD^{-1}=\ee^{2\rb{\gconf-\fconf}}\sfBBS,
\end{alignat}
\end{subequations}
We note the following property of the elementary symmetric polynomials,
\begin{align}
	e_n(fX)	&=(fX)^{[a_1}{}_{a_1}(fX)^{a_2}{}_{a_2}\cdots (fX)^{a_n]}{}_{a_n} \nonumber \\
				&= \rb{f^n} X^{[a_1}{}_{a_1}X^{a_2}{}_{a_2}\cdots X^{a_n]}{}_{a_n}=f^ne_n(X),
\end{align}
where $f$ is a scalar field. We use this property to compute the bimetric interactions in terms of the BSSN variables,
\begin{subequations}
\label{eq:biminteractionsBSSN2}
\begin{align}
	\sgV	&\coloneqq -\betaSum e_n(\sfD) =-\betaSum e_n(\ee^{2\rb{\fconf-\gconf}}\sfDBS)=-\betaSum \ee^{2n\rb{\fconf-\gconf}}e_n(\sfDBS) \\
	\sfV	&\coloneqq -\sLt^{-1}\betaSum e_{n-1}(\sgB) =-\sLt^{-1}\betaSum e_{n-1}(\ee^{2\rb{\fconf-\gconf}}\sgBBS) \nonumber \\
			&=-\sLt^{-1}\betaSum \ee^{2\rb{n-1}\rb{\fconf-\gconf}}e_{n-1}(\sgBBS) \\
	\sgU 	&\coloneqq-\sLt^{-1}\betaSum Y_{n-1}(\sgB) = -\sLt^{-1}\betaSum\sum_{k=0}^{n-1}(-1)^{n-1+k}e_k(\sgB)\sgB^{n-1-k} \nonumber \\
			&=-\sLt^{-1}\betaSum\sum_{k=0}^{n-1}(-1)^{n-1+k}e_k(\ee^{2\rb{\fconf-\gconf}}\sgBBS)\ee^{2\rb{n-1-k}\rb{\fconf-\gconf}}\sgBBS^{n-1-k} \nonumber \\
			&=-\sLt^{-1}\betaSum\sum_{k=0}^{n-1}(-1)^{n-1+k}e_k(\sgBBS)\ee^{2k\rb{\fconf-\gconf}}\ee^{2\rb{n-1-k}\rb{\fconf-\gconf}}\sgBBS^{n-1-k} \nonumber \\
			&=-\sLt^{-1}\betaSum\ee^{2\rb{n-1}\rb{\fconf-\gconf}}\sum_{k=0}^{n-1}(-1)^{n-1+k}e_k(\sgBBS)\sgBBS^{n-1-k} \nonumber \\
			&=-\sLt^{-1}\betaSum\ee^{2\rb{n-1}\rb{\fconf-\gconf}}Y_{n-1}(\sgBBS),\label{eq:sgUBS} \\
	\sfU	&\coloneqq -\sfD\betaSum Y_{n-1}(\sfD)=-\sfD\betaSum\sum_{k=0}^{n-1}(-1)^{n-1+k}e_k(\sfD)\sfD^{n-1-k} \nonumber \\
	&=-\ee^{2\rb{\fconf-\gconf}}\sfDBS\betaSum\sum_{k=0}^{n-1}(-1)^{n-1+k}e_k(\ee^{2\rb{\fconf-\gconf}}\sfDBS)\ee^{2\rb{n-1-k}\rb{\fconf-\gconf}}\sfDBS^{n-1-k} \nonumber \\
	&=-\ee^{2\rb{\fconf-\gconf}}\sfDBS\betaSum\sum_{k=0}^{n-1}(-1)^{n-1+k}e_k(\sfDBS)\ee^{2k\rb{\fconf-\gconf}}\ee^{2\rb{n-1-k}\rb{\fconf-\gconf}}\sfDBS^{n-1-k} \nonumber \\
	&=-\sfDBS\betaSum\ee^{2n\rb{\fconf-\gconf}}\sum_{k=0}^{n-1}(-1)^{n-1+k}e_k(\sfDBS)\sfDBS^{n-1-k} \nonumber \\
	&=-\sfDBS\betaSum\ee^{2n\rb{\fconf-\gconf}}Y_{n-1}(\sfDBS). \label{eq:sfUBS}
\end{align}
\end{subequations}
From \mref{eq:sQU,eq:sQBS,eq:sgUBS,eq:sfUBS} it follows,
\begin{align}
	\sgQU	&=-\sgQBS\sfDBS\betaSum\ee^{2n\rb{\fconf-\gconf}}Y_{n-1}(\sfDBS) =-\sgBBS\betaSum\ee^{2n\rb{\fconf-\gconf}}Y_{n-1}(\sfDBS), \\
	\sfQU	&=-\sLt^{-1}\sfQBS\betaSum\ee^{2\rb{n-1}\rb{\fconf-\gconf}}Y_{n-1}(\sgBBS).
\end{align}

We compute the bimetric sources in \eqref{eq:bimexplicitsources} in terms of the BSSN variables,
\begin{subequations}
\label{eq:bimsourcesBSSN}
\begin{align}
	\grhob		&\coloneqq \betaSum e_n(\sgB) =\betaSum e_n(\ee^{2\rb{\fconf-\gconf}}\sgBBS)=\betaSum \ee^{2n\rb{\fconf-\gconf}}e_n(\sgBBS), \\
	\frhob		&\coloneqq \sLt\betaSum e_{n-1}(\sfD)\mydet{\gE\fE^{-1}}= \sLt\betaSum e_{n-1}(\ee^{2\rb{\fconf-\gconf}}\sfDBS)\mydet{\ee^{2\rb{\gconf-\fconf}}\gEBS\fEBS^{-1}} \nonumber \\
					&= \sLt\betaSum \ee^{2\rb{n-1}\rb{\fconf-\gconf}}\ee^{2\sdim\rb{\gconf-\fconf}}e_{n-1}(\sfDBS)\mydet{\gEBS\fEBS^{-1}} \nonumber \\
					&= \sLt\betaSum \ee^{2\rb{n-1-\sdim}\rb{\fconf-\gconf}}e_{n-1}(\sfDBS)\mydet{\gEBS\fEBS^{-1}}, \\
	\gjotab		&\coloneqq \betaSum \gSp\sgQU\sgn = -\betaSum \ee^{4\gconf}\gSpBS\sgBBS\ee^{2n\rb{\fconf-\gconf}}Y_{n-1}(\sfDBS) \ee^{-2\gconf}\sgnBS  \nonumber \\
					&= -\betaSum \ee^{2n\fconf-2(n-1)\gconf} \gSpBS\sgBBS Y_{n-1}(\sfDBS)\sgnBS,  \\
	\fjotab		&\coloneqq - \gjotab \mydet{\gE\fE^{-1}} = \mydet{\ee^{2\rb{\gconf-\fconf}}\gEBS\fEBS^{-1}}\betaSum \ee^{2n\fconf-2(n-1)\gconf}\gSpBS\sgBBS Y_{n-1}(\sfDBS)\sgnBS  \nonumber \\
					&= \mydet{\gEBS\fEBS^{-1}}\betaSum \ee^{2\sdim\rb{\gconf-\fconf}}\ee^{2n\fconf-2(n-1)\gconf}\gSpBS\sgBBS Y_{n-1}(\sfDBS)\sgnBS  \nonumber \\
					&= \mydet{\gEBS\fEBS^{-1}}\betaSum \ee^{2\rb{n-\sdim}\fconf-2(n-1-\sdim)\gconf}\gSpBS\sgBBS Y_{n-1}(\sfDBS)\sgnBS,  \\
	\gJotab		&\coloneqq \rb{\sgV I-\sgQU+\mW\sgU} \nonumber \\
					&=-\betaSum\rb{\ee^{2n\rb{\fconf-\gconf}}e_n(\sfDBS) I -\ee^{2n\rb{\fconf-\gconf}}\sgBBS Y_{n-1}(\sfDBS)+\mW \sLt^{-1}\ee^{2\rb{n-1}\rb{\fconf-\gconf}}Y_{n-1}(\sgBBS)} \nonumber \\
					&=-\betaSum\ee^{2n\rb{\fconf-\gconf}}\rb{e_n(\sfDBS) I -\sgBBS Y_{n-1}(\sfDBS)+\mW \sLt^{-1}\ee^{2\rb{\gconf-\fconf}}Y_{n-1}(\sgBBS)}, \\
	\fJotab		&\coloneqq \rb{\sfV I-\sfQU+\mW^{-1}\sfU}\mydet{\gE\fE^{-1}} \nonumber \\
					&=-\betaSum\left(\sLt^{-1} \ee^{2\rb{n-1}\rb{\fconf-\gconf}}e_{n-1}(\sgBBS) I -\sLt^{-1}\sfQBS\ee^{2\rb{n-1}\rb{\fconf-\gconf}}Y_{n-1}(\sgBBS)\right. \nonumber \\
					&\qquad\qquad\qquad\;\left.+\mW^{-1}\sfDBS\ee^{2n\rb{\fconf-\gconf}}Y_{n-1}(\sfDBS) \right)\mydet{\ee^{2\rb{\gconf-\fconf}}\gEBS\fEBS^{-1}} \nonumber \\
					&=-\betaSum\ee^{2\rb{n-\sdim}\rb{\fconf-\gconf}}\left(\sLt^{-1} \ee^{2\rb{\gconf-\fconf}}e_{n-1}(\sgBBS) I-\sLt^{-1}\sfQBS\ee^{2\rb{\gconf-\fconf}}Y_{n-1}(\sgBBS)\right. \nonumber \\
					&\qquad\qquad\qquad\qquad\qquad\quad\;\,\left.+\mW^{-1}\sfDBS Y_{n-1}(\sfDBS)\vphantom{\ee^{2\rb{\gconf-\fconf}}}\right)\mydet{\gEBS\fEBS^{-1}}.
\end{align}
\end{subequations}
We remind the reader that $\sdim=\stdim-1$ is the dimension of the spacelike hypersurface, in our case $\sdim=3$, and $I$ is the $\sdim\times \sdim$ identity. Note that both the bimetric interactions and sources can be computed starting with the physical vielbeins and metrics $\gE,\fE,\gSp,\fSp$, or with the conformal vielbeins and metrics $\gEBS,\fEBS,\gSpBS,\fSpBS$ and the conformal factors $\gconf,\fconf$.

\subsection{The bimetric standard \nPlusOne equations}
\label{app:stdeqs}

The bimetric standard \nPlusOne evolution equations read \cite{Kocic:2018ddp},
\begin{subequations}
\label{eq:BEE}
	\begin{alignat}{3}
		\pfg \gSp_{ij}		&=-2\gLapse \gK_{ij}, && & \\
		\pfg \gK_{ij}		&=-\gD_i\gD_j \gLapse &&+\gLapse\qb{\gR_{ij}-2\gK_{ik}\gK{}^k{}_j+\gK\gK_{ij}}& \nonumber \\
								& &&-\gLapse \gKappa\qb{\gSp_{ik}\gJotaeff{}^k{}_j-\dfrac{1}{\sdim-1}\gSp_{ij}\rb{\gJotaeff^i{}_i-\grhoeff}}, \\
		\pff \fSp_{ij}		&=-2\fLapse \fK_{ij}, && & \\
		\pff \fK_{ij}		&=-\fD_i\fD_j \fLapse &&+\fLapse\qb{\fR_{ij}-2\fK_{ik}\fK{}^k{}_j+\fK\fK_{ij}}& \nonumber \\
								& &&-\fLapse \fKappa\qb{\fSp_{ik}\fJotaeff{}^k{}_j-\dfrac{1}{\sdim-1}\fSp_{ij}\rb{\fJotaeff^i{}_i-\frhoeff}},
	\end{alignat}
\end{subequations}
with $\pfg\coloneqq \partial_t-\mathscr{L}_\gShift $, $\pff\coloneqq \partial_t-\mathscr{L}_\fShift$, $\gShift$ and $\fShift$ being the shift vectors of the two metrics.
The bimetric standard \nPlusOne constraint equations are,
\begin{subequations}
\label{eq:BCE}
	\begin{alignat}{3}
		2\gCC		&\coloneqq \gR+\gK^2-\gK_{ij}\gK^{ij}-2\gKappa \grhoeff	\ &=0, \\
		\gCC_i		&\coloneqq \gD_k\gK^k{}_i-\gD_i\gK-\gKappa\gjotaeff{}_i		&=0, \\
		2\fCC			&\coloneqq \fR+\fK^2-\fK_{ij}\fK^{ij}-2\fKappa \frhoeff			&=0, \\
		\fCC_i		&\coloneqq \fD_k\fK^k{}_i-\fD_i\fK-\fKappa\fjotaeff{}_i			&=0.
	\end{alignat}
\end{subequations}
The effective sources are the sum of the bimetric sources given by \eqref{eq:bimexplicitsources}, and the external matter sources,
\begin{subequations}
\label{eq:effsources}
	\begin{align}
		\grhoeff		&= \grhob + \grho, \qquad \gjotaeff_i= \gjotab_i +\gjota_i, \qquad \gJotaeff_{ij}= \gJotab_{ij} +\gJota_{ij}, \\
		\frhoeff		&= \frhob + \frho, \qquad \fjotaeff_i= \fjotab_i +\fjota_i, \qquad \fJotaeff_{ij}= \fJotab_{ij} +\fJota_{ij}.
	\end{align}
\end{subequations}
Note that the relation between the two bimetric currents in \eqref{eq:bimexplicitsources}, implies that the two momentum constraints are related to each other,\footnote{Note, however, that solving one momentum constraint does not imply that the other is automatically satisfied. Both of them need to be solved independently.}
\begin{align}
\label{eq:MCrelation}
	\sqrt{\gSp}\cb{\gKappa^{-1}\rb{\gD_k\gK^k{}_i-\gD_i\gK}-\gjota_i}+\sqrt{\fSp}\cb{ \fKappa^{-1}\rb{\fD_k\fK^k{}_i-\fD_i\fK}-\fjota_i	}=0.
\end{align}
The bimetric conservation law (BCL), in its asymmetric and symmetric form, reads
\begin{subequations}
\label{eq:SC}
	\begin{alignat}{3}
			\label{eq:SCa}
	\mS	&\coloneqq\sgU^i{}_j \rb{\gD_i\sgn^j-\gK^j{}_i}+\sfU^i{}_j \rb{\fD_i\sfn^j+\fK^j{}_i}-\gD_i\rb{\sgU^i{}_j\sgn^j} \ 				& \\
			&=\dfrac{1}{2}\gD_i\rb{\sgU^i{}_j\sgn^j}+\dfrac{1}{2}\fD_i\rb{\sfU^i{}_j\sfn^j}																& \nonumber	\\
			\label{eq:SCb}
			&\quad -\sgU^i{}_j\rb{\gD_i\sgn^j-\dfrac{1}{2}\dfrac{\partial_i\sqrt{\gdet}}{\sqrt{\gdet}}\sgn^j-\gK^j{}_i}-\sfU^i{}_j\rb{\fD_i\sfn^j-\dfrac{1}{2}\dfrac{\partial_i\sqrt{\fdet}}{\sqrt{\fdet}}\sfn^j+\fK^j{}_i}			&=0.
\end{alignat}
\end{subequations}

\subsection{The bimetric BSSN equations}
\label{app:BSSNeqs}

We now write down the bimetric BSSN equations, applying the procedure in \cite[Sec.~11.5]{baumgarte2010numerical} to \mref{eq:BEE,eq:BCE,eq:SC}, using \eqref{eq:bimsourcesBSSN} for the bimetric sources.

The bimetric BSSN \threePlusOne evolution equations for the $g$-sector are,
\begin{subequations}
\label{eq:BSSNEEg}
	\begin{align}
		\pfg \gconf 	&=-\dfrac{\gLapse\gK}{6}, \\
		\pfg \gK	&=-\gSp^{ij}\gD_i\gD_j\gLapse+\gLapse \qb{\gABS^{ij}\gABS_{ij}+\dfrac{\gK^2}{3}+\dfrac{\gKappa}{2} \rb{\gJotaeff^i{}_i+\grhoeff}}, \\
		\pfg \gSpBS_{ij}	&=-2\gLapse \gABS_{ij}, \\
		\pfg \gABS_{ij}	&=\ee^{-4\gconf}\cb{-\gD_i\gD_j\gLapse+\dfrac{1}{3}\gSp_{ij}\gD_k\gD^k\gLapse+\gLapse \qb{\gR_{ij}-\dfrac{1}{3}\gSp_{ij}\gR-\gKappa \rb{\gJotaeff_{ij}-\dfrac{1}{3}\gSp_{ij}\gJotaeff^i{}_i}}}, \\
		\pfg \gGBS^i	&=-2\gABS^{ij}\partial_j \gLapse +2\gLapse \qb{\gGBS^i_{jk}\gABS^{kj}-\dfrac{2}{3}\gSpBS^{ij}\partial_j\gK+6\gABS^{ij}\partial_j\gconf-\gKappa\gSpBS^{ij}\gjotaeff_j},
	\end{align}
\end{subequations}
where
\begin{subequations}
\label{eq:Liedersg}
	\begin{align}
		\mathscr{L}_\gShift \gconf	&=\gShift^i\partial_i \gconf +\dfrac{1}{6}\partial_k\gShift^k, \qquad\mathscr{L}_\gShift \gK	=\gShift^i\partial_i\gK, \\
		\mathscr{L}_\gShift \gSpBS_{ij}	&=\gShift^k\partial_k \gSpBS_{ij}+\gSpBS_{ik}\partial_j\gShift^k+\gSpBS_{jk}\partial_i\gShift^k-\dfrac{2}{3}\gSpBS_{ij}\partial_k\gShift^k, \\
		\mathscr{L}_\gShift \gABS_{ij}		&=\gShift^k\partial_k \gABS_{ij}+\gABS_{ik}\partial_j\gShift^k+\gABS_{jk}\partial_i\gShift^k-\dfrac{2}{3}\gABS_{ij}\partial_k\gShift^k, \\
		\mathscr{L}_\gShift \gGBS^i		&=\gShift^j\partial_j\gGBS^i-\gGBS^j\partial_j\gShift^i+\dfrac{2}{3}\gGBS^i\partial_j\gShift^j+\dfrac{1}{3}\gSpBS^{ki}\partial_k\partial_j\gShift^j+\gSpBS^{kj}\partial_k\partial_j\gShift^i,
	\end{align}
\end{subequations}
and for the $f$-sector
\begin{subequations}
\label{eq:BSSNEEf}
	\begin{align}
		\pff \fconf 	&=-\dfrac{\fLapse\fK}{6}, \\
		\pff \fK	&=-\fSp^{ij}\fD_i\fD_j\fLapse+\fLapse \qb{\fABS^{ij}\fABS_{ij}+\dfrac{\fK^2}{3}+\dfrac{\fKappa}{2} \rb{\fJotaeff^i{}_i+\frhoeff}}, \\
		\pff \fSpBS_{ij}	&=-2\fLapse \fABS_{ij}, \\
		\pff \fABS_{ij}	&=\ee^{-4\fconf}\cb{-\fD_i\fD_j\fLapse+\dfrac{1}{3}\fSp_{ij}\fD_k\fD^k\fLapse+\fLapse \qb{\fR_{ij}-\dfrac{1}{3}\fSp_{ij}\fR-\fKappa \rb{\fJotaeff_{ij}-\dfrac{1}{3}\fSp_{ij}\fJotaeff^i{}_i}}}, \\
		\pff \fGBS^i	&=-2\fABS^{ij}\partial_j \fLapse +2\fLapse \qb{\fGBS^i_{jk}\fABS^{kj}-\dfrac{2}{3}\fSpBS^{ij}\partial_j\fK+6\fABS^{ij}\partial_j\fconf-\fKappa\fSpBS^{ij}\fjotaeff_j},
	\end{align}
\end{subequations}
where
\begin{subequations}
\label{eq:Liedersf}
	\begin{align}
		\mathscr{L}_\fShift \fconf	&=\fShift^i\partial_i \fconf +\dfrac{1}{6}\partial_k\fShift^k, \qquad\mathscr{L}_\fShift \fK	=\fShift^i\partial_i\fK, \\
		\mathscr{L}_\fShift \fSpBS_{ij}	&=\fShift^k\partial_k \fSpBS_{ij}+\fSpBS_{ik}\partial_j\fShift^k+\fSpBS_{jk}\partial_i\fShift^k-\dfrac{2}{3}\fSpBS_{ij}\partial_k\fShift^k, \\
		\mathscr{L}_\fShift \fABS_{ij}		&=\fShift^k\partial_k \fABS_{ij}+\fABS_{ik}\partial_j\fShift^k+\fABS_{jk}\partial_i\fShift^k-\dfrac{2}{3}\fABS_{ij}\partial_k\fShift^k, \\
		\mathscr{L}_\fShift \fGBS^i		&=\fShift^j\partial_j\fGBS^i-\fGBS^j\partial_j\fShift^i+\dfrac{2}{3}\fGBS^i\partial_j\fShift^j+\dfrac{1}{3}\fSpBS^{ki}\partial_k\partial_j\fShift^j+\fSpBS^{kj}\partial_k\partial_j\fShift^i.
	\end{align}
\end{subequations}
The bimetric BSSN constraint equations are
\begin{subequations}
\label{eq:BSSNBCE}
	\begin{alignat}{3}
		\gCCBS			&\coloneqq \gSpBS^{ij}\gDBS_i\gDBS_j \ee^{\gconf}-\dfrac{\ee^{\gconf}}{8}\gR+\dfrac{\ee^{5\gconf}}{8}\gABS^{ij}\gABS_{ij}-\dfrac{\ee^{5\gconf}}{12}\gK^2+\dfrac{\gKappa}{4}\ee^{5\gconf}\grhoeff \ \ &=0, \\
		\gCCBS_i		&\coloneqq \gDBS_j\rb{\ee^{6\gconf}\gABS^{ij}}-\dfrac{2}{3}\gSpBS^{ij}\gDBS_{j}\gK-\gKappa\ee^{6\gconf}\gjotaeff^i	 &=0, \\
		\gCGBS			&\coloneqq \gGBS^i+\partial_j\gSpBS^{ij}	 &=0, \\
		\fCCBS			&\coloneqq \fSpBS^{ij}\fDBS_i\fDBS_j \ee^{\fconf}-\dfrac{\ee^{\fconf}}{8}\fR+\dfrac{\ee^{5\fconf}}{8}\fABS^{ij}\fABS_{ij}-\dfrac{\ee^{5\fconf}}{12}\fK^2+\dfrac{\fKappa}{4}\ee^{5\fconf}\frhoeff			&=0, \\
		\fCCBS_i			&\coloneqq \fDBS_j\rb{\ee^{6\fconf}\fABS^{ij}}-\dfrac{2}{3}\fSpBS^{ij}\fDBS_{j}\fK-\fKappa\ee^{6\fconf}\fjotaeff^i				&=0, \\
		\fCGBS			&\coloneqq	\fGBS^i+\partial_j\fSpBS^{ij}	 &=0.
	\end{alignat}
\end{subequations}

Now we compute the BSSN decomposition of the bimetric conservation law \eqref{eq:SC}. We note that, due to \eqref{eq:confmetrics}, the relation between the two conformally related covariant derivatives of a vector field $X^i$ is,
\begin{align}
	\gD_i X^j =\gDBS_i X^j+2\rb{X^j\partial_i\gconf+\delta_i{}^jX^k\partial_k\gconf-\gSpBS_{i\ell} X^\ell\gSpBS^{j k}\partial_k\gconf},
\end{align}
which implies
\begin{align}
	\gD_i X^i =\gDBS_i X^i+6X^i\partial_i \gconf,
\end{align}
with analogous formulas for the $\fMet$-sector. Using the definition of the conformal extrinsic curvatures \eqref{eq:confAcov} in the cBSSN formalism, the bimetric conservation law becomes
\begin{subequations}
\label{eq:SCBSSN}
	\begin{alignat}{3}
	\mS	&\coloneqq\sgU^i{}_j \qb{\gDBS_i \sgn^j+2\rb{\sgn^j\partial_i\gconf+\delta_i{}^j\sgn^k\partial_k\gconf-\gSpBS_{i\ell} \sgn^\ell\gSpBS^{j k}\partial_k\gconf}	-\ee^{4\gconf}\rb{\gABS^j{}_i+\dfrac{1}{3}\delta^j{}_i\gKBS}} \nonumber \\
			&\quad+\sfU^i{}_j \qb{\fDBS_i \sfn^j+2\rb{\sfn^j\partial_i\fconf+\delta_i{}^j\sfn^k\partial_k\fconf-\fSpBS_{i\ell} \sfn^\ell\fSpBS^{j k}\partial_k\fconf}	+\ee^{4\fconf}\rb{\fABS^j{}_i+\dfrac{1}{3}\delta^j{}_i\fKBS}}\nonumber \\
			&\quad-\gDBS_i\rb{\sgU^i{}_j\sgn^j}-6\sgU^i{}_j\sgn^j\partial_i\gconf \ 				& \label{eq:SCBSSNa}\\
			&=\dfrac{1}{2}\qb{\gDBS_i\rb{\sgU^i{}_j\sgn^j}-6\sgU^i{}_j\sgn^j\partial_i\gconf}+\dfrac{1}{2}\dfrac{\partial_i\sqrt{\gdet}}{\sqrt{\gdet}}\sgU^i{}_j\sgn^j\nonumber \\
			&\quad+\dfrac{1}{2}\qb{\fDBS_i\rb{\sfU^i{}_j\sfn^j}-6\sfU^i{}_j\sfn^j\partial_i\fconf}-	\dfrac{1}{2}\dfrac{\partial_i\sqrt{\fdet}}{\sqrt{\fdet}}\sfU^i{}_j\sfn^j															& \nonumber	\\
			&\quad -\sgU^i{}_j\qb{\gDBS_i \sgn^j+2\rb{\sgn^j\partial_i\gconf+\delta_i{}^j\sgn^k\partial_k\gconf-\gSpBS_{i\ell} \sgn^\ell\gSpBS^{j k}\partial_k\gconf}-\ee^{4\gconf}\rb{\gABS^j{}_i+\dfrac{1}{3}\delta^j{}_i\gKBS}} \nonumber \\
			&\quad-\sfU^i{}_j\qb{\fDBS_i \sfn^j+2\rb{\sfn^j\partial_i\fconf+\delta_i{}^j\sfn^k\partial_k\fconf-\fSpBS_{i\ell} \sfn^\ell\fSpBS^{j k}\partial_k\fconf}+\ee^{4\fconf}\rb{\fABS^j{}_i+\dfrac{1}{3}\delta^j{}_i\fKBS}}			=0.\label{eq:SCBSSNb}&
\end{alignat}
\end{subequations}
Finally, we insert the expressions computed in Appendix~\ref{app:BSSNbimint} for the quantities appearing in \eqref{eq:SCBSSN}, which are written in terms of the BSSN variables.

\subsection{The bimetric covariant BSSN equations}
\label{app:cBSSNeqs}

At this point, we apply the method described in \cite{PhysRevD.79.104029} to the bimetric BSSN equations in Appendix~\ref{app:BSSNeqs}. The bimetric interactions and sources are not changed.

These are the cBSSN \threePlusOne bimetric evolution equations for the $g$-sector, assuming that the background connections do not depend on time,
\begin{subequations}
\label{eq:cBSSNEEg}
	\begin{align}
		\pfg \gconf 			&=-\dfrac{\pfg \log \rb{\gdetBS}}{12}-\dfrac{\gLapse}{6}\rb{\gABS+\gKBS}, \\
		\pfg \gKBS				&=-\pfg\gABS+\dfrac{\gLapse}{3}\rb{\gKBS^2+2\gKBS\gABS}+\gLapse\gABS_{ij}\gABS^{ij} && \nonumber \\
									&\quad\hspace{0.35mm} -\ee^{4\gconf}\rb{\gDBS_i\gDBS^i\gLapse+2\gDBS^i\gLapse\gDBS_i\gconf}+\dfrac{\gKappa}{2}\gLapse \rb{\gJotaeff^i{}_i+\grhoeff}, \\
		\pfg \gSpBS_{ij}		&=\dfrac{1}{3}\gSpBS_{ij}\pfg \log \rb{\gdetBS}-2\gLapse \rb{\gABS_{ij}-\dfrac{1}{3}\gSpBS_{ij}\gABS}, && \\
		\pfg \gABS_{ij}		&=\dfrac{1}{3}\gABS_{ij}\pfg \log \rb{\gdetBS}+\dfrac{1}{3}\gABS_{ij}\pfg \gABS && \nonumber \\
									&\quad\hspace{0.35mm} +\gLapse\cb{-2\gABS_{ik}\gABS^{k}{}_j+\gKBS\gABS_{ij}+\dfrac{1}{3}\gABS\qb{5\gABS_{ij}-\rb{\gKBS+\gABS}\gSpBS_{ij}}} \nonumber \\
									&\quad\hspace{0.35mm} +\ee^{-4\gconf}\cb{-\gDBS_i\gDBS_j\gLapse+4\gDBS_{(i}\gLapse\gDBS_{j)}\gconf+\gLapse \qb{\gRBS_{ij}-\gDBS_i\gDBS_j\gconf+4\gDBS_{i}\gconf\gDBS_{j}\gconf-\gKappa \gJotaeff_{ij}}}^\mathrm{TF}, \\
		\pfg \gL^i				&=\gSpBS^{jk}\gDback_j\gDback_k\gShift^i-\gSpBS^{jk}\gRback^{i}_{jk\ell}\gShift^\ell \nonumber \\
									&\quad\hspace{0.35mm} -\dfrac{1}{3}\gL^i\pfg\log\rb{\gdet}-\dfrac{1}{6}\gSpBS^{ij}\partial_j\pfg\log\rb{\gdet}-\dfrac{4}{3}\gLapse \gSp^{ij}\partial_j\gKBS \nonumber \\
									&\quad\hspace{0.35mm} -2 \rb{\gABS^{jk}-\dfrac{1}{3}\gSpBS^{jk}\gABS}\rb{\delta^i{}_j\partial_k\gLapse-6\gLapse\delta^i{}_j\partial_k\gconf-\gLapse \gDG^i_{jk}}-2\gKappa\gLapse\ee^{4\gconf}\gjotaeff^i,\label{eq:cBSSNEEgL}
	\end{align}
\end{subequations}
where $\gDback$ and $\gRback$ are the covariant derivative and the Riemann tensor of the background geometry, and
\begin{subequations}
\label{eq:cLiedersg}
	\begin{align}
		\mathscr{L}_\gShift \gconf			&=\gShift^i\partial_i \gconf , \qquad\mathscr{L}_\gShift \gK	=\gShift^i\partial_i\gK, \\
		\mathscr{L}_\gShift \gSpBS_{ij}	&=\gShift^k\partial_k \gSpBS_{ij}+\gSpBS_{ik}\partial_j\gShift^k+\gSpBS_{jk}\partial_i\gShift^k, \\
		\mathscr{L}_\gShift \gABS_{ij}		&=\gShift^k\partial_k \gABS_{ij}+\gABS_{ik}\partial_j\gShift^k+\gABS_{jk}\partial_i\gShift^k, \\
		\mathscr{L}_\gShift \gGBS^i		&=\gShift^j\partial_j\gGBS^i-\gGBS^j\partial_j\gShift^i, \\
		\gRBS_{ij}									&\coloneqq -\dfrac{1}{2}\gSpBS^{k\ell}\gDback_{k}\gDback_{\ell}\gSpBS_{ij}+\gSpBS_{k(i}\gDback_{j)}\gL^k-\gSpBS^{k\ell}\gSpBS_{m(i}\gRback_{j)k\ell}{}^m \nonumber \\
														&\quad\ +\gSpBS^{\ell m}\gDG^k_{\ell m}\gDG_{(ij)k}+\gSpBS^{k\ell}\rb{2\gDG^m_{k(i}\gDG_{j)m\ell}+\gDG^m_{ik}\gDG_{mj\ell}}.\label{eq:Ricciidentity}
	\end{align}
\end{subequations}
Note that all Lie derivatives in \eqref{eq:cLiedersg} are Lie derivatives of tensors, not tensor densities as in \eqref{eq:Liedersg}. Equation \eqref{eq:Ricciidentity} is an identity for the Ricci tensor of $\gSpBS_{ij}$ in terms of the background geometry \cite{PhysRevD.79.104029,PhysRevD.85.084004} (see Appendix A of \cite{PhysRevD.85.084004} for the proof). The superscript $\mathrm{TF}$ means \qm{trace-free}. Note that all the traces are with respect to $\gSpBS$ except the trace of $\gJotaeff$, which is with respect to $\gSp$. For the $\fMet$-sector we have,
\begin{subequations}
\label{eq:cBSSNEEf}
	\begin{align}
		\pff \fconf 				&=-\dfrac{\pff \ln \rb{\fdetBS}}{12}-\dfrac{\fLapse}{6}\rb{\fABS+\fKBS}, \\
		\pff \fKBS				&=-\pff\fABS+\dfrac{\fLapse}{3}\rb{\fKBS^2+2\fKBS\fABS}+\fLapse\fABS_{ij}\fABS^{ij} \nonumber \\
									&\quad\hspace{0.35mm}-\ee^{4\fconf}\rb{\fDBS_i\fDBS^i\fLapse+2\fDBS^i\fLapse\fDBS_i\fconf}+\dfrac{\fKappa}{2}\fLapse \rb{\fJotaeff^i{}_i+\frhoeff}, \\
		\pff \fSpBS_{ij}		&=\dfrac{1}{3}\fSpBS_{ij}\pff \log \rb{\fdetBS}-2\fLapse \rb{\fABS_{ij}-\dfrac{1}{3}\fSpBS_{ij}\fABS}, \\
		\pff \fABS_{ij}			&=\dfrac{1}{3}\fABS_{ij}\pff \log \rb{\fdetBS}+\dfrac{1}{3}\fABS_{ij}\pff \fABS \nonumber \\
									&\quad\hspace{0.35mm}+\fLapse\cb{-2\fABS_{ik}\fABS^{k}{}_j+\fKBS\fABS_{ij}+\dfrac{1}{3}\fABS\qb{5\fABS_{ij}-\rb{\fKBS+\fABS}\fSpBS_{ij}}} \nonumber \\
									&\quad\hspace{0.35mm}+\ee^{-4\fconf}\cb{-\fDBS_i\fDBS_j\fLapse+4\fDBS_{(i}\fLapse\fDBS_{j)}\fconf+\fLapse \qb{\fRBS_{ij}-\fDBS_i\fDBS_j\fconf+4\fDBS_{i}\fconf\fDBS_{j}\fconf-\fKappa \fJotaeff_{ij}}}^\mathrm{TF}, \\
		\pff \fL^i				&=\fSpBS^{jk}\fDback_j\fDback_k \fShift^i -\fSpBS^{jk}\fRback^{i}_{jk\ell}\fShift^\ell \nonumber \\
									&\quad\hspace{0.35mm}-\dfrac{1}{3}\fL^i\pff\log\rb{\fdet}-\dfrac{1}{6}\fSpBS^{ij}\partial_j\pff\log\rb{\fdet}-\dfrac{4}{3}\fLapse \fSp^{ij}\partial_j\fKBS \nonumber \\
									&\quad\hspace{0.35mm}-2 \rb{\fABS^{jk}-\dfrac{1}{3}\fSpBS^{jk}\fABS}\rb{\delta^i{}_j\partial_k\fLapse-6\fLapse\delta^i{}_j\partial_k\fconf-\fLapse \fDG^i_{jk}}-2\fLapse\ee^{4\fconf}\fKappa\fjotaeff^i,
	\end{align}
\end{subequations}
where the same conventions and notations as for the $\gMet$-sector are implied. It holds,
\begin{subequations}
\label{eq:cLiedersf}
	\begin{align}
		\mathscr{L}_\fShift \fconf			&=\fShift^i\partial_i \fconf , \qquad\mathscr{L}_\fShift \fK	=\fShift^i\partial_i\fK, \\
		\mathscr{L}_\fShift \fSpBS_{ij}		&=\fShift^k\partial_k \fSpBS_{ij}+\fSpBS_{ik}\partial_j\fShift^k+\fSpBS_{jk}\partial_i\fShift^k, \\
		\mathscr{L}_\fShift \fABS_{ij}		&=\fShift^k\partial_k \fABS_{ij}+\fABS_{ik}\partial_j\fShift^k+\fABS_{jk}\partial_i\fShift^k, \\
		\mathscr{L}_\fShift \fL^i				&=\fShift^j\partial_j\fGBS^i-\fGBS^j\partial_j\fShift^i+\dfrac{2}{3}\fGBS^i\partial_j\fShift^j, \\
		\fRBS_{ij}									&\coloneqq -\dfrac{1}{2}\fSpBS^{k\ell}\fDback_{k}\fDback_{\ell}\fSpBS_{ij}+\fSpBS_{k(i}\fDback_{j)}\fL^k-\fSpBS^{k\ell}\fSpBS_{m(i}\fRback_{j)k\ell}{}^m \nonumber \\
														&\quad\ +\fSpBS^{\ell m}\fDG^k_{\ell m}\fDG_{(ij)k}+\fSpBS^{k\ell}\rb{2\fDG^m_{k(i}\fDG_{j)m\ell}+\fDG^m_{ik}\fDG_{mj\ell}}.
	\end{align}
\end{subequations}
The bimetric cBSSN constraint equations are
\begin{subequations}
\label{eq:cBSSNBCE}
	\begin{alignat}{3}
		\gCCBS			&\coloneqq \dfrac{2}{3}\rb{\gKBS+\gABS}^2+\dfrac{1}{3}\gABS^2- \gABS_{ij}\gABS^{ij} \ \ & \nonumber \\
							&\quad\; +\ee^{-4\gconf}\rb{\gRiBS-8\gDBS_i\gconf\gDBS^i\gconf-8\gDBS^i\gDBS_i\gconf}-2\gKappa\grhoeff \ \ &=0, \\
		\gCCBS^i		&\coloneqq \ee^{-4\gconf}\left\lbrace\dfrac{1}{\sqrt{\gdet}}\gDback_j\rb{\sqrt{\gdet}\gABS^{ij}}+6\rb{\gABS^{ij}-\dfrac{1}{3}\gSpBS^{ij}\gABS}\partial_j\gconf\right.	 & \nonumber \\
							&\left.\quad\qquad\quad\,-\gSpBS^{ij}\partial_j\rb{\dfrac{2}{3}\gKBS+\gABS}+\gABS^{jk}\gDG^i_{jk}\right\rbrace-\gKappa \gjotaeff^i		 &=0, \\
		\gCGBS			&\coloneqq \gL^i-\gSpBS^{jk}\rb{\gGBS ^i_{jk}-\gGbackBS ^i_{jk}} &=0, \\
		\fCCBS			&\coloneqq \dfrac{2}{3}\rb{\fKBS+\fABS}^2+\dfrac{1}{3}\fABS^2- \fABS_{ij}\fABS^{ij} \ \ & \nonumber \\
							&\quad\; +\ee^{-4\fconf}\rb{\fRiBS-8\fDBS_i\fconf\fDBS^i\fconf-8\fDBS^i\fDBS_i\fconf}-2\fKappa\frhoeff  \ \	&=0, \\
		\fCCBS^i		&\coloneqq \ee^{-4\fconf}\left\lbrace\dfrac{1}{\sqrt{\fdet}}\fDback_j\rb{\sqrt{\fdet}\fABS^{ij}}+6\rb{\fABS^{ij}-\dfrac{1}{3}\fSpBS^{ij}\fABS}\partial_j\fconf\right.	 & \nonumber \\
							&\left.\quad\qquad\quad\,-\fSpBS^{ij}\partial_j\rb{\dfrac{2}{3}\fKBS+\fABS}+\fABS^{jk}\fDG^i_{jk}\right\rbrace-\fKappa \fjotaeff^i		 &=0, \\
		\fCGBS			&\coloneqq \fL^i-\fSpBS^{jk}\rb{\fGBS ^i_{jk}-\fGbackBS ^i_{jk}}	 &=0.
	\end{alignat}
\end{subequations}
The cBSSN BCL is the same as in one in \eqref{eq:SCBSSN}, since the bimetric interactions \mref{eq:biminteractionsBSSN,eq:biminteractionsBSSN2} are not changed, and the bimetric sources are the same as in \eqref{eq:bimsourcesBSSN}.

\subsection{The bimetric covariant BSSN equations in spherical symmetry}
\label{app:sphsym}

The equations in this Appendix were computed with the Mathematica package $\mathtt{bimEX}$ \cite{TORSELLO2020106948}, which can compute the bimetric cBSSN equations for any desired ansatz.

We write down the cBSSN equations in spherical symmetry, assuming the following ansatz,
\begin{alignat}{4}
	\gEBS^\textbf{a}{}_i 	&= \diagmat{\gA (t,r),\gB (t,r) ,\gB (t,r)\sin (\theta)},& \; \fEtrBS^\textbf{a}{}_i 	&= \diagmat{\fA (t,r),\fB (t,r) ,\fB (t,r)\sin (\theta)},& \nonumber \\
	\gABS^i{}_j 	&= \diagmat{\gAo (t,r),\gAt (t,r) ,\gAt (t,r)},&  \fABS^i{}_j 	&= \diagmat{\fAo (t,r),\fAt (t,r) ,\fAt (t,r)},& \nonumber \\
	\gL^i	 	&= \mat{\gLo (t,r) \\ 0 \\ 0}, \quad  \fL^i= \mat{\fLo (t,r) \\ 0 \\ 0},& \quad \hShift ^i &= \mat{\qo (t,r) \\ \qt (t,r) \\ \qth (t,r)}, \quad \sLp^\textbf{a}= \mat{\po (t,r) \\ \pt (t,r) \\ \pth (t,r)}. &
\end{alignat}
The background geometries for both $\gSp$ and $\fSp$ are chosen to be the spatial part of the flat metric in spherical coordinates,\footnote{Note that this is not the Minkowski metric in the Lorentz frame, whose spatial part remains $\sEta=\diagmat{1,1,1}$.}
\begin{align}
	\delta = \diagmat{1,r^2,r^2\sin(\theta)^2}.
\end{align}
From now on, we will assume the time and radial dependence of all the fields. We define the function,
\begin{align}
	\mR 	&\coloneqq \dfrac{\ee^{2\fconf}\fB}{\ee^{2\gconf}\gB},
\end{align}
and the \qm{shifted elementary symmetric polynomials} \cite{doi:10.1063/1.5100027},
\begin{align}
	\esp{n}{k}	&\coloneqq \sum_{i=0}^n \binom{n}{i} \beta_{(i+k)} \mR^i, \quad \binom{n}{i} = \dfrac{n!}{i!(n-i)!},
\end{align}
where the $\beta_{(n)}$ are five real dimensionless constants appearing in the bimetric interaction potential \cite{Hassan:2011zd}.

Consistency conditions between the various equations imply that $\qt=\pt=0$, and $\qth,\pth$ do not appear explicitly into the equations, so we can set them to zero without losing generality. Also, $\sRs$ is the identity and $\fEBS=\fEtrBS$, which simplifies the computations considerably.

The constraint equations \eqref{eq:cBSSNBCE} read,
\begin{subequations}
\label{eq:cBSSNBCEsph}
	\begin{alignat}{3}
		\gCCBS			&= \dfrac{2}{\gA^2\ee^{4\gconf}}\left\lbrace\dfrac{\gA^2}{\gB^2}+\rb{\dfrac{2\dr\gA}{\gA}-\dfrac{\dr\gB}{\gB}-2\dr\gconf}\rb{\dfrac{\dr\gB}{\gB}+2\dr\gconf}-2\rb{\dfrac{\drr \gB}{\gB}+2\drr \gconf +2\dfrac{\dr\gB}{\gB}\dr\gconf}\right\rbrace \ \ & \nonumber \\
							&\quad+\dfrac{2}{3}\rb{\gKBS+2\gAo+\gAt}\rb{\gKBS+3\gAt}-2\gKappa \rb{\grho+\grhob}  =0,& \\
		\gCCBS^r		&=\dfrac{\ee^{-4\gconf}}{\gA^2}\left\lbrace 2\rb{\dfrac{\dr\gB}{\gB}+2\dr\gconf}\rb{\gAo-\gAt}-2\dr\rb{\gAt+\dfrac{1}{3}\gKBS}\right\rbrace -\gKappa\rb{\gjota^r +\gjotab^r}=0,&\label{eq:cBSSNBCEsphMCg} \\
		\gCGBS			&= \gLo -\rb{\dfrac{2r}{\gB^2}+\dfrac{\dr\gA}{\gA^3}-\dfrac{2\dr\gB}{\gA^2\gB}}=0,& \\
		\fCCBS			&= \dfrac{2}{\fA^2\ee^{4\fconf}}\left\lbrace\dfrac{\fA^2}{\fB^2}+\rb{\dfrac{2\dr\fA}{\fA}-\dfrac{\dr\fB}{\fB}-2\dr\fconf}\!\!\rb{\dfrac{\dr\fB}{\fB}+2\dr\fconf}\!-2\rb{\dfrac{\drr \fB}{\fB}+2\drr \fconf +2\dfrac{\dr\fB}{\fB}\dr\fconf}\right\rbrace \ \ & \nonumber \\
							&\quad+\dfrac{2}{3}\rb{\fKBS+2\fAo+\fAt}\rb{\fKBS+3\fAt}-2\fKappa \rb{\frho+\frhob}  =0,& \\
		\fCCBS^r		&=\dfrac{\ee^{-4\fconf}}{\fA^2}\left\lbrace 2\rb{\dfrac{\dr\fB}{\fB}+2\dr\fconf}\rb{\fAo-\fAt}-2\dr\rb{\fAt+\dfrac{1}{3}\fKBS}\right\rbrace -\fKappa\rb{\fjota^r +\fjotab^r}=0,&\label{eq:cBSSNBCEsphMCf} \\
		\fCGBS			&= \fLo -\rb{\dfrac{2r}{\fB^2}+\dfrac{\dr\fA}{\fA^3}-\dfrac{2\dr\fB}{\fA^2\fB}} =0,&
	\end{alignat}
\end{subequations}
where the bimetric energy density and currents are given by
\begin{subequations}
\label{eq:bimendenssph}
	\begin{alignat}{3}
		\grhob	&= -\esp{2}{0}-\dfrac{\fA\ee^{2\fconf}}{\gA\ee^{2\gconf}}\esp{2}{1}\,\sLt, \qquad &\gjotab^r&=-\dfrac{\fA\ee^{2\fconf}}{\gA^2\ee^{4\gconf}}\esp{2}{1}\,\po, \\
		\frhob	&= -\dfrac{1}{\mR^2}\rb{\dfrac{\esp{2}{2}}{\mR^2}+\dfrac{\gA\ee^{2\gconf}}{\fA\ee^{2\fconf}}\esp{2}{1}\,\sLt}, \qquad &\fjotab^r&=\dfrac{\gA\ee^{2\gconf}}{\fA^2\ee^{4\fconf}\mR^2}\esp{2}{1}\,\po= -\dfrac{\gA^3\ee^{6\gconf}}{\fA^3\ee^{6\fconf}\mR^2}\, \gjotab^r .
	\end{alignat}
\end{subequations}
We can solve $\po$ from both \eqref{eq:cBSSNBCEsphMCg} and \eqref{eq:cBSSNBCEsphMCf}, obtaining respectively
\begin{subequations}
\label{eq:psph}
	\begin{align}
		\po	&=\dfrac{\ee^{-2\fconf}}{\fA\esp{2}{1}}\cb{\ee^{4\gconf}\gA^2\gjota^r+\dfrac{2}{\gKappa}\qb{\dr\rb{\gAt+\dfrac{1}{3}\gKBS}-\rb{2\dr\gconf+\dfrac{\dr\gB}{\gB}}\rb{\gAo-\gAt}}}, \\
		\po	&=-\dfrac{\ee^{-2\gconf}\mR^2}{\gA\esp{2}{1}}\cb{\ee^{4\fconf}\fA^2\fjota^r+\dfrac{2}{\fKappa}\qb{\dr\rb{\fAt+\dfrac{1}{3}\fKBS}-\rb{2\dr\fconf+\dfrac{\dr\fB}{\fB}}\rb{\fAo-\fAt}}}.
	\end{align}
\end{subequations}
The asymmetric---and simpler---version of the BCL \eqref{eq:SCBSSNa} is,
\begin{alignat}{3}
\label{eq:bimconssphsym}
	\mS	&= \dfrac{\dr\po\esp{2}{1}}{\sLt} +2\sLt\qb{\ee^{2\gconf}\gA\rb{\fAt+\dfrac{1}{3}\fKBS}\mR\esp{1}{1}-\ee^{2\fconf}\fA\rb{\gAt+\dfrac{1}{3}\gKBS}\esp{1}{2}} \nonumber \\
			&\quad +2\,\dfrac{\ee^{2\gconf}\gA}{\ee^{2\fconf}\fA}\,\po\!\qb{\!\rb{\dfrac{\dr\gB}{\gB}\!+\!2\dr\gconf}\!\!\!\rb{\mR\esp{1}{1}\!+\!\dfrac{\ee^{4\fconf}\fA^2}{\ee^{4\gconf}\gA^2}\esp{1}{2}}\!+\!\dfrac{\ee^{2\gconf}}{\ee^{2\fconf}}\rb{\mR\,\dr\rb{\gconf\!-\!\fconf}\!+\!\dfrac{\dr\mR}{2}}\!}\!\esp{1}{1} \nonumber \\
			&\quad - \ee^{2\fconf}\fA\qb{-\rb{\fKBS+\fABS}\esp{2}{1}+2\rb{\fAt+\dfrac{1}{3}\fKBS}\esp{1}{1}} \nonumber \\
			&\quad -\ee^{2\gconf}\gA\qb{\rb{\gAo+\dfrac{1}{3}\gKBS}\esp{2}{1}+2\rb{\gAt+\dfrac{1}{3}\gKBS}\esp{1}{1}}	=0.&
\end{alignat}

The evolution equations \mref{eq:cBSSNEEg,eq:cBSSNEEf}, modified to get the evolution of the components $\gABS^i{}_j,\fABS^i{}_j$ rather than $\gABS_{ij},\fABS_{ij}$, reduce to
\begin{subequations}
\label{eq:cBSSNEEsph}
	\begin{align}
		\dt \gconf 				&=\gShift^r\dr\gconf-\dfrac{\pfg \log \rb{\gdetBS}}{12}-\dfrac{\gLapse}{6}\rb{\gABS+\gKBS}, \\
		\dt \gKBS				&=\gShift^r\dr\gKBS-\pfg\gABS+\dfrac{\gLapse}{3}\rb{\gKBS^2+2\gKBS\gABS}+\gLapse\rb{\gAo^2+2\gAt^2} \nonumber \\
									&\quad+\dfrac{\ee^{4\gconf}\dr\gLapse}{\gA^2}\rb{\dfrac{\dr\gA}{\gA}-2\dr\gconf-\dfrac{\dr\gB}{\gB}}+\dfrac{\gKappa}{2}\gLapse \rb{\gJota^i{}_i+\grho+\gJotab^i{}_i+\grhob}, \\
		\dt \gA					&=\gShift^r\dr\gA+\gA\qb{\dr\gShift^r+\dfrac{1}{6}\pfg \log \rb{\gdetBS}-\gLapse \rb{\gAo-\dfrac{1}{3}\gABS}}, \\
		\dt \gB					&=\gShift^r\dr\gB+\gB\qb{\dfrac{1}{6}\pfg \log \rb{\gdetBS}-\gLapse \rb{\gAt-\dfrac{1}{3}\gABS}}, \\
		\dt \gAo					&=\gShift^r\dr\gAo+\dfrac{\pfg\gABS}{3}+\gLapse\rb{\gABS+\gKBS}\!\rb{\gAo-\dfrac{\gABS}{3}}-\dfrac{2\gKappa}{3}\gLapse\rb{\gJota^r{}_r-\gJota^\theta{}_\theta+\gJotab^r{}_r-\gJotab^\theta{}_\theta}  \nonumber \\
									&\quad+\dfrac{2\ee^{-4\gconf}}{3\gA^2}\left\lbrace  \gLapse\qb{2\dr\gconf\rb{\dfrac{\dr\gA}{\gA}+2\dr\gconf+\dfrac{\dr\gB}{\gB}}-2\drr\gconf+\gRBS_{11}-\dfrac{\gA^2}{\gB^2}\gRBS_{22}} \right. \nonumber \\		
									&\qquad\qquad\quad\;\;\left.+\dr\gLapse \rb{\dfrac{\dr\gA}{\gA} +4\dr\gconf+\dfrac{\dr\gB}{\gB}}-\drr\gLapse \right\rbrace \\
		\dt \gAt					&=\gShift^r\dr\gAt+\dfrac{\pfg\gABS}{3}+\gLapse\rb{\gABS+\gKBS}\!\rb{\gAt-\dfrac{\gABS}{3}}+\dfrac{\gKappa}{3}\gLapse\rb{\gJota^r{}_r-\gJota^\theta{}_\theta+\gJotab^r{}_r-\gJotab^\theta{}_\theta} \nonumber \\
									&\quad-\dfrac{\ee^{-4\gconf}}{3\gA^2}\left\lbrace  \gLapse\qb{2\dr\gconf\rb{\dfrac{\dr\gA}{\gA}+2\dr\gconf+\dfrac{\dr\gB}{\gB}}-2\drr\gconf+\gRBS_{11}-\dfrac{\gA^2}{\gB^2}\gRBS_{22}} \right. \nonumber \\		
									&\qquad\qquad\quad\left. +\dr\gLapse \rb{\dfrac{\dr\gA}{\gA} +4\dr\gconf+\dfrac{\dr\gB}{\gB}}-\drr\gLapse 	 \right\rbrace	\\
		\dt \gLo			&=\gShift^r\rb{\dr\gLo-\dfrac{2}{\gB^2}}+\dfrac{\drr\gShift^r}{\gA^2}-\dfrac{4\dr\gLapse}{3\gA^2}\rb{\gAo-\gAt}-2\gKappa\ee^{4\gconf}\gLapse\rb{\gjota^r+\gjotab^r} \nonumber \\
									&\quad +\dfrac{1}{6\gA^2}\qb{\pfg\log(\gdet)\dr\log(\gdet)-\dr\pfg\log(\gdet)}-\dfrac{\gLo}{3}\pfg\log(\gdet)+\dr\gShift^r\rb{\dfrac{2r}{\gB^2}-\gLo} \nonumber \\
									&\quad +\dfrac{4\gLapse}{3}\qb{\dfrac{\rb{\gAo-\gAt}}{\gA^2}\rb{\dfrac{\dr\gA}{\gA}+6\dr\gconf+\dfrac{\dr\gB}{\gB}}-\dfrac{\rb{\gAo-\gAt}r}{\gB^2}-\dfrac{\dr\rb{\gABS+\gKBS}}{\gA^2}}, \\
	\dt \fconf 				&=\fShift^r\dr\fconf-\dfrac{\pff \log \rb{\fdetBS}}{12}-\dfrac{\fLapse}{6}\rb{\fABS+\fKBS}, \\
		\dt \fKBS				&=\fShift^r\dr\fKBS-\pff\fABS+\dfrac{\fLapse}{3}\rb{\fKBS^2+2\fKBS\fABS}+\fLapse\rb{\fAo^2+2\fAt^2} \nonumber \\
									&\quad+\dfrac{\ee^{4\fconf}\dr\fLapse}{\fA^2}\rb{\dfrac{\dr\fA}{\fA}-2\dr\fconf-\dfrac{\dr\fB}{\fB}}+\dfrac{\fKappa}{2}\fLapse \rb{\fJota^i{}_i+\frho+\fJotab^i{}_i+\frhob}, \\
		\dt \fA					&=\fShift^r\dr\fA+\fA\qb{\dr\fShift^r+\dfrac{1}{6}\pff \log \rb{\fdetBS}-\fLapse \rb{\fAo-\dfrac{1}{3}\fABS}}, \\
		\dt \fAo					&=\fShift^r\dr\fAo+\dfrac{\pff\fABS}{3}+\fLapse\rb{\fABS+\fKBS}\!\rb{\fAo-\dfrac{\fABS}{3}}-\dfrac{2\fKappa}{3}\fLapse\rb{\fJota^r{}_r-\fJota^\theta{}_\theta+\fJotab^r{}_r-\fJotab^\theta{}_\theta}  \nonumber \\
									&\quad+\dfrac{2\ee^{-4\fconf}}{3\fA^2}\left\lbrace  \fLapse\qb{2\dr\fconf\rb{\dfrac{\dr\fA}{\fA}+2\dr\fconf+\dfrac{\dr\fB}{\fB}}-2\drr\fconf+\fRBS_{11}-\dfrac{\fA^2}{\fB^2}\fRBS_{22}} \right. \nonumber \\		
									&\qquad\qquad\quad\;\;\left.+\dr\fLapse \rb{\dfrac{\dr\fA}{\fA} +4\dr\fconf+\dfrac{\dr\fB}{\fB}}-\drr\fLapse \right\rbrace \\
		\dt \fAt					&=\fShift^r\dr\fAt+\dfrac{\pff\fABS}{3}+\fLapse\rb{\fABS+\fKBS}\!\rb{\fAt-\dfrac{\fABS}{3}}+\dfrac{\fKappa}{3}\fLapse\rb{\fJota^r{}_r-\fJota^\theta{}_\theta+\fJotab^r{}_r-\fJotab^\theta{}_\theta} \nonumber \\
									&\quad-\dfrac{\ee^{-4\fconf}}{3\fA^2}\left\lbrace  \fLapse\qb{2\dr\fconf\rb{\dfrac{\dr\fA}{\fA}+2\dr\fconf+\dfrac{\dr\fB}{\fB}}-2\drr\fconf+\fRBS_{11}-\dfrac{\fA^2}{\fB^2}\fRBS_{22}} \right. \nonumber \\		
									&\qquad\qquad\quad\left. +\dr\fLapse \rb{\dfrac{\dr\fA}{\fA} +4\dr\fconf+\dfrac{\dr\fB}{\fB}}-\drr\fLapse 	 \right\rbrace	\\
		\dt \fLo			&=\fShift^r\rb{\dr\fLo-\dfrac{2}{\fB^2}}+\dfrac{\drr\fShift^r}{\fA^2}+\dfrac{4\dr\fLapse}{3\fA^2}\rb{\fAo-\fAt}-2\fKappa\ee^{4\fconf}\fLapse\rb{\fjota^r+\fjotab^r} \nonumber \\
									&\quad +\dfrac{1}{6\fA^2}\qb{\pff\log(\fdet)\dr\log(\fdet)-\dr\pff\log(\fdet)}-\dfrac{\fLo}{3}\pff\log(\fdet)+\dr\fShift^r\rb{\dfrac{2r}{\fB^2}-\fLo} \nonumber \\
									&\quad +\dfrac{4\fLapse}{3}\qb{\dfrac{\rb{\fAo-\fAt}}{\fA^2}\rb{\dfrac{\dr\fA}{\fA}+6\dr\fconf+\dfrac{\dr\fB}{\fB}}-\dfrac{\rb{\fAo-\fAt}r}{\fB^2}-\dfrac{\dr\rb{\fABS+\fKBS}}{\fA^2}}
	\end{align}
\end{subequations}
where
\begin{subequations}
\label{eq:Ricciidentitysph}
	\begin{align}
		\gRBS_{11}	&= \dfrac{3(\dr\gA)^2}{\gA^2}-\dfrac{\drr\gA}{\gA}+\gA^2\rb{\dr\gLo-\dfrac{2}{\gB^2}}-\dfrac{2(\dr\gB)^2}{\gB^2}-\dfrac{2\dr\gA\dr\gB}{\gA\gB}+\dfrac{4r\gA^2\dr\gB}{\gB^3}, \\
		\gRBS_{22}	&= -1-\dfrac{(\dr\gB)^2}{\gA^2}-\dfrac{\gB\drr\gB}{\gA^2}+\dr\gA\rb{\dfrac{\gB\dr\gB}{\gA^3}-\dfrac{\gB^2}{\gA^3r}}+\dfrac{2\gB\dr\gB}{\gA^2r}+\dfrac{\gB^2\gLo}{r}, \\
		\fRBS_{11}	&= \dfrac{3(\dr\fA)^2}{\fA^2}-\dfrac{\drr\fA}{\fA}+\fA^2\rb{\dr\fLo-\dfrac{2}{\fB^2}}-\dfrac{2(\dr\fB)^2}{\fB^2}-\dfrac{2\dr\fA\dr\fB}{\fA\fB}+\dfrac{4r\fA^2\dr\fB}{\fB^3}, \\
		\fRBS_{22}	&= -1-\dfrac{(\dr\fB)^2}{\fA^2}-\dfrac{\fB\drr\fB}{\fA^2}+\dr\fA\rb{\dfrac{\fB\dr\fB}{\fA^3}-\dfrac{\fB^2}{\fA^3r}}+\dfrac{2\fB\dr\fB}{\fA^2r}+\dfrac{\fB^2\fLo}{r},
	\end{align}
\end{subequations}
and
\begin{subequations}
\label{eq:bimsourcessph}
	\begin{align}
		\gJotab^r{}_r 	&=\esp{2}{0}+\dfrac{\mW\esp{2}{1}}{\sLt}-\ee^{2\rb{\fconf-\gconf}}\dfrac{\fA\po^2\esp{2}{1}}{\gA\sLt}, \\
		\gJotab^\theta{}_\theta 	&=\esp{1}{0}+\dfrac{\mW\esp{1}{1}}{\sLt}+\ee^{2\rb{\fconf-\gconf}}\rb{\dfrac{\fA\esp{1}{1}}{\gA\sLt}+\dfrac{\fA\mW\esp{1}{2}}{\gA}}, \\
		\gJotab^i{}_i 	&=\gJota^r{}_r+2\gJota^\theta{}_\theta =2\esp{1}{0}+\esp{2}{0}+\dfrac{\mW}{\sLt}\rb{2\esp{1}{1}+\esp{2}{1}} \nonumber \\
							&\qquad\qquad\qquad\quad+\ee^{2\rb{\fconf-\gconf}}\rb{-\dfrac{\sLt\fA\esp{2}{1}}{\gA}+\dfrac{\fA\rb{2\esp{1}{1}+\esp{2}{1}}}{\gA\sLt}+\dfrac{2\fA\mW\esp{1}{2}}{\gA}}, \\
		\fJotab^r{}_r 	&=\dfrac{\esp{2}{2}}{\mR^2}+\dfrac{\esp{2}{1}}{\mR^2\mW\sLt}-\ee^{2\rb{\gconf-\fconf}}\dfrac{\gA\po^2\esp{2}{1}}{\mR^2\fA\sLt}, \\
		\fJotab^\theta{}_\theta 	&=\dfrac{\esp{2}{2}-\esp{1}{2}}{\mR^2}+\dfrac{\esp{1}{2}}{\mR\mW\sLt}+\ee^{2\rb{\gconf-\fconf}}\qb{\dfrac{\gA\rb{\esp{2}{1}-\esp{1}{1}}}{\mR^2\fA\sLt}+\dfrac{\gA\esp{1}{1}}{\mR\mW\fA}}, \\
		\fJotab^i{}_i 	&=\fJota^r{}_r+2\fJota^\theta{}_\theta =\dfrac{3\esp{2}{2}-2\esp{1}{2}}{\mR^2}+\dfrac{1}{\mW\mR^2\sLt}\rb{2\mR\esp{1}{2}+\esp{2}{1}} \nonumber \\
							&\qquad\qquad\qquad\quad+\ee^{2\rb{\gconf-\fconf}}\rb{-\dfrac{\sLt\gA\esp{2}{1}}{\mR^2\fA}+\dfrac{\gA\rb{3\esp{2}{1}-2\esp{1}{1}}}{\mR^2\fA\sLt}+\dfrac{2\gA\esp{1}{1}}{\mW\mR\fA}}.		
	\end{align}
\end{subequations}
The evolution equation for $\po$ is obtained from the cBSSN version of (A.8) in \cite{Kocic:2018ddp},
\begin{align}
\label{eq:dtp}
	\esp{2}{1}\dt\po	&=\esp{2}{1}\qo\dr\po + \esp{2}{1}\sLt \rb{\dfrac{\dr\fLapse}{\ee^{2\fconf}\fA}-\dfrac{\dr\gLapse}{\ee^{2\gconf}\gA}} \nonumber \\
								&\quad +\gLapse\,\po\qb{\esp{2}{1}\rb{\gAo+\dfrac{1}{3}\gKBS}+2\rb{\gAt+\dfrac{1}{3}\gKBS}\esp{1}{1}} \nonumber\\
								 &\quad +\fLapse\,\po\qb{\esp{2}{1}\rb{\fAo+\dfrac{1}{3}\fKBS}+2\rb{\fAt+\dfrac{1}{3}\fKBS}\mR\esp{1}{2}} \nonumber\\
								&\quad +\dfrac{2}{\ee^{2\gconf}\gB}\rb{\gLapse\esp{1}{1}\dfrac{\dr\fB+2\,\fB\,\dr\fconf}{\fA}-\fLapse\esp{1}{2}\dfrac{\dr\gB+2\,\gB\,\dr\gconf}{\gA}} \nonumber \\
								&\quad +\dfrac{2\sLt}{\ee^{2\gconf}\gB}\rb{
								     \fLapse\esp{1}{2}\dfrac{\dr\fB+2\,\fB\,\dr\fconf}{\fA}-\gLapse\esp{1}{1}\dfrac{\dr\gB+2\,\gB\,\dr\gconf}{\gA}}.
\end{align}
A direct comparison between \eqref{eq:dtp} and (2.10) in \cite{Kocic_2019}, reveals their equivalence.

Equations \mref{eq:cBSSNBCEsph,eq:bimendenssph,eq:psph,eq:bimconssphsym,
eq:cBSSNEEsph,eq:Ricciidentitysph,eq:bimsourcessph,eq:dtp} reduce to the standard \threePlusOne equations in spherical symmetry presented in \cite{Kocic:2018ddp} after imposing,
\begin{alignat}{4}
	\gKBS	 &=0, 	\qquad	& \fKBS &=0, \qquad & \gABS &=\gAo+2\gAt, \qquad & \quad\fABS &=\fAo+2\fAt, \nonumber \\
	\gABS_i		&=\gK _i,	& \fABS_i   &=\fK_i, & \gLo &=\rb{\dfrac{2r}{\gB^2}+\dfrac{\dr\gA}{\gA^3}-\dfrac{2\dr\gB}{\gA^2\gB}}, &\fLo &=\rb{\dfrac{2r}{\fB^2}+\dfrac{\dr\fA}{\fA^3}-\dfrac{2\dr\fB}{\fA^2\fB}}, \nonumber \\
	\gconf &=0, & \fconf &=0, & \pfg \log(\gdet) &=-2\gLapse \rb{\gK _1+2\gK _2}, & \pff \log(\fdet) &= -2\fLapse \rb{\fK _1+2\fK _2},
\end{alignat}
and appropriately using the Hamiltonian constraints in the evolution equations for the traces $\gKBS$ and $\fKBS$, and the momentum constraints in the evolution equations for the conformal factors $\gconf,\fconf$.

\bibliographystyle{JHEP}
\bibliography{biblio}

\end{document}